\documentclass[journal]{IEEEtran}

\usepackage{setspace}
\ifCLASSOPTIONonecolumn
\fi

\usepackage[top=2.0cm, bottom=2.2cm, left=2.2cm,right=2.2cm]{geometry}
\IEEEoverridecommandlockouts
\usepackage{cite}
\usepackage{amsmath,amssymb,amsfonts}
\usepackage{adjustbox}
\usepackage{graphicx}
\usepackage{booktabs}
\usepackage{multirow}
\usepackage{placeins}
\usepackage{textcomp}
\usepackage{color}
\usepackage{comment}
\usepackage{psfig}
\usepackage{multirow}
\usepackage{steinmetz} 
\usepackage{tabularx}
\usepackage[multiple]{footmisc} 
\newcommand{\mycomment}[1]{%
}%
\usepackage{environ}
\NewEnviron{nestedcomment}{\mycomment{\BODY}}

\usepackage{algorithm}
\usepackage{algpseudocode}

\usepackage[font=small,labelfont=bf]{caption}

\usepackage{etoolbox}
\makeatletter
\patchcmd{\@makecaption}
  {\scshape}
  {}
  {}
  {}
\makeatother

\def\urltilde{\kern -.15em\lower .7ex\hbox{\~{}}\kern .04em}

\DeclareMathOperator{\sinc}{sinc}
\allowdisplaybreaks
\begin{document}
\title{Discrete Beamforming Optimization for RISs with a Limited Phase Range and Amplitude Attenuation
\thanks{D. K. Pekcan and E. Ayanoglu are, and H. Liao was with the Department of Electrical Engineering and Computer Science,
University of California, Irvine. H. Liao will be with the Shanghai Research Institute for Autonomous Systems, Tongji University, China.}
\thanks{This work is partially supported by NSF grant 2030029.}
}

\ifCLASSOPTIONonecolumn
\author{\IEEEauthorblockN{Dogan Kutay Pekcan, {\em Graduate Student Member, IEEE}}\\
\IEEEauthorblockN{Hongyi Liao, {\em Graduate Student Member, IEEE}}\\
\and
\IEEEauthorblockN{Ender Ayanoglu, {\em Fellow, IEEE}}\\
}
\else
\author{\IEEEauthorblockN{Dogan Kutay Pekcan, {\em Graduate Student Member, IEEE}}\\
\IEEEauthorblockN{Hongyi Liao, {\em Graduate Student Member, IEEE}}\\
\IEEEauthorblockN{Ender Ayanoglu, {\em Fellow, IEEE}}\\
}
\fi

\maketitle

\begin{abstract}
This paper addresses the problem of maximizing the received power at a user equipment via reconfigurable intelligent surface (RIS) characterized by phase-dependent amplitude (PDA) and discrete phase shifts over a limited phase range. 
Given complex RIS coefficients, that is, discrete phase shifts and PDAs, we derive the necessary and sufficient conditions to achieve the optimal solution.
To this end, we propose an optimal search algorithm that is proven to converge in linear time within at most $NK$ steps, significantly outperforming the exhaustive search approach that would otherwise be needed for RISs with amplitude attenuation. 
Furthermore, we introduce a practical quantization framework for PDA-introduced RISs termed \textit{amplitude-introduced polar quantization} (APQ), and extend it to a novel algorithm named \textit{extended amplitude-introduced polar quantization} (EAPQ) that works with geometric projections. We derive closed-form expressions to assess how closely the performance of the proposed RIS configuration can approximate the ideal case with continuous phases and no attenuation. 
Our analysis reveals that increasing the number of discrete phases beyond $K=4$ yields only marginal gains, regardless of attenuation levels, provided the RIS has a sufficiently wide phase range $R$. 
Furthermore, we also show and quantify that when the phase range $R$ is limited, the performance is sensitive to attenuation for larger $R$, and sensitive to $R$ when there is less attenuation.
Finally, the proposed optimal algorithm provides a generic upper bound that could serve as a benchmark for discrete beamforming in RISs with amplitude constraints.
\end{abstract}

\section{Problem Definition} \label{sec:ProblemDefinition}
%

\begin{figure}[!t]
\centering
\resizebox{0.8\linewidth}{!}{\input{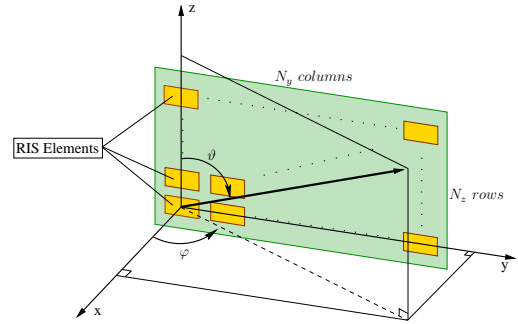}}
\caption{RIS structure.}
\label{fig:ris}
\end{figure}

%
Consider a Reconfigurable Intelligent Surface (RIS) structure as in Fig.~\ref{fig:ris}. This structure is intended to be used as a reflector for transmissions from a base station (BS) to a user equipment (UE) when the direct link from the BS to the UE may be blocked. Let the combination of the BS-to-RIS and the RIS-to-UE channels for the $n$-th RIS element be $h_n,$ $n=1, 2, \ldots, N$ where $N$ is the number of individual RIS elements. We assume $h_n = \beta_n e^{j \alpha_n}$ with $\beta_n \geq 0$ and $\alpha \in [-\pi, \pi),$ and the $n$-th RIS element introduces a complex number $\beta^r(\theta_n) e^{j\theta_n}$ to $h_n$ with $\beta^r(\theta_n) \in [0,1]$ and $\theta_n$ taking values from a discrete phase shift set $\Phi_K = \{\phi_1, \phi_2, \dots, \phi_K\}$ where $K$ is an integer, $n=1, 2, \ldots, N$. We let $h_0 = \beta_0e^{j\alpha_0}$ be the direct link between the BS and UE. We state that $\beta^r(\theta_n): [-\pi, \pi] \rightarrow [0,1]$ is a function of $\theta_n$ representing the gain of the $n$-th RIS element. 
We refer to this dependence between the RIS phase shift $\theta_n$ and the respective gain $\beta^r(\theta_n)$ for the $n$-th RIS element as the phase dependent amplitude (PDA) model. To accommodate for the discrete phase shift constraint and the PDA model, we define ${\bf W}_K$, where ${\bf W}_K = \{\beta^r(\phi_1)e^{j\phi_1}, \beta^r(\phi_2)e^{j\phi_2}, \ldots, \beta^r(\phi_K)e^{j\phi_K}\}$, using the phases in $\Phi_K$ together with the respective gains. Therefore, RIS coefficients will belong to this coefficients set, i.e., ${\rm w}_n \in {\bf W}_K$ for $n=1,\ldots,N$. We also define the difference among each adjacent phase shift in $\Phi_K$ as $\Omega_K=\{\omega_1, \omega_2, \dots, \omega_K\}$, such that $\phi_{k\oplus1} = \phi_k + \omega_k$.\footnote{In this paper, we define $\oplus$ and $\ominus$ to choose from RIS phase shift indexes from $1$ to $K$ as follows. 
For $k_1,k_2 \in \{1,\dots,K\}$, $k_1 \oplus k_2 = k_1 + k_2$ if $k_1+k_2 \leq K$ and $k_1 \oplus k_2 = k_1+k_2-K$, otherwise. Similarly, for $k_1,k_2 \in \{1,\dots,K\}$, $k_1 \ominus k_2 = k_1 - k_2$ if $k_1 > k_2$ and $k_1 \ominus k_2 = K + k_1 - k_2$, otherwise.}

Our goal is to maximize the received power at the UE, that is, $| h_0 + \sum_{n=1}^N h_n \beta^r(\theta_n) e^{j\theta_n} |^2,$ which can be formally described as
\begin{equation}
\begin{aligned}
 & \underset{\mbox{\boldmath$\theta$}}{\rm maximize\ } f({\mbox{\boldmath$\theta$}})\\
 & {\rm subject\ to\ } \theta_n\in \Phi_K,\ n=1, 2, \ldots, N
\end{aligned}
\label{eqn:problem1_pda}
\end{equation}
where
\begin{equation}\label{eq:finitial_theta_pda}
f({\mbox{\boldmath$\theta$}}) = \bigg|\beta_0e^{j\alpha_0}+\sum_{n=1}^N \beta_n
\beta^r(\theta_n) e^{j(\alpha_n + \theta_n)}\bigg|^2,
\end{equation}
$\beta_n \geq 0,\ n=0,1,\dots,N$, ${\mbox{\boldmath$\theta$}} = (\theta_1, \theta_2, \ldots, \theta_N)$, and $\alpha_n \in [-\pi,\pi)$ for $n=0,1,\dots,N$.

The objective term $f({\mbox{\boldmath$\theta$}})$ can alternatively be written as $f_1({\bf w}) = | h_0 + \sum_{n=1}^N h_n {\rm w}_n |^2$ and is therefore connected to the generic $K$-ary discrete quadratic program (QP). In \cite{PA24}, the globally optimum solution for this problem was achieved in the least number of steps for a uniform discrete phase shift set with unit RIS gains, i.e., $\omega_k = 2\pi/K, k=1, \ldots, K,$ and $\beta^r(\theta_n)=1$, for $\theta_n \in [-\pi, \pi)$. 
In \cite{PLA24}, the problem was addressed for a nonuniform discrete phase shift set with adjustable RIS gains, i.e., $\beta^r \in [0,1]$ being arbitrarily adjustable. In this paper, we provide optimal and suboptimal but effective algorithms for this problem, where there is a dependence between the phase shift selection of the RIS element and its gain. 

In the following sections, first, we define the PDA model for the discrete phase shift structures given an RIS phase range. 
Second, we provide necessary and sufficient conditions to achieve the globally optimum solution with discrete phase shifts based on arbirary ${\bf W}_K$. 
Then, we approach this problem intuitively and come up with low-complexity quantization algorithms. 
We quantify the performance of the globally optimum algorithm and those of the low-complexity quantization algorithms and show that they are close.
Finally, we analyze the RIS capability in terms of approximating the ideal scenario for large $N$, where the ideal scenario corresponds to using continuous phases with no amplitude attenuation in the RIS gains.

\section{Phase Dependent Amplitude and the RIS Phase Range}\label{sec:PDAandPHASERANGE_MODEL}

Practical RISs have certain limitations due to hardware constraints that do not favor the use of continuous phase shifts or the assumption of unit gains across all elements. Also, they allow only a certain range of phases to be used \cite{PYTCLWZB21, AZWY20}. 
First, we observe that it is difficult to realize continuous phase shifts, and control bits should be used to implement a finite set of phase shifts for practicality. 
Second, in practice, RISs can provide a certain range of phase shifts that potentially results in a nonuniform structure for the set of discrete phase shifts \cite{PYTCLWZB21}. 
Finally, due to hardware limitations, the reflection amplitudes of the RIS are not necessarily constant over all the elements, and actually depend on the selected phase shift for each element \cite{AZWY20}.

In this paper, besides providing a globally optimum algorithm for the problem in (\ref{eqn:problem1_pda}), our aim is also to quantify the expected performance of the RIS in terms of how well the maximum achievable power can be approximated. The maximum achievable power corresponds to the ideal scenario, similar to \cite{PA24, PLA24}, where the practical concerns, such as the PDA model or the use of discrete phase shifts are not taken into account. Building on this ideality, we focus on the following practical constraints:
\begin{itemize}
    \item Using discrete phase shifts that are within the RIS phase range.
    \item Considering variable gains over RIS elements that depend on the selected phase shift, i.e., the PDA model.
\end{itemize}

We will first explain the discrete phase shift structure. Then, we will refer to the closed form equation from the literature for a practical RIS gain model.

\subsection{The RIS Phase Range and the Discrete Phase Shifts}

The RIS phase range $R \in [0, 2\pi]$ represents the phase-shifting capability of the RIS, allowing phase shift selections that are in the range $[-R/2, R/2],$ without loss of generality \cite{PLA24, PYTCLWZB21}. 
The ideal scenario plays a significant role in this paper especially to quantify the performance loss arising from the non-ideal constraints, e.g., the derivation of the approximation ratio is based on the ideal scenario.
Therefore, when it comes to placing the discrete phase shifts, we also consider the ideal scenario, where we make the discrete phase shifts equally separated within the RIS phase range based on the proven optimality in \cite{PLA24}. This also helps present structured numerical results that are simple to follow.
\ifCLASSOPTIONonecolumn
\begin{figure*}[!t]
\centering
\begin{minipage}{0.48\textwidth}
\centering
\includegraphics[width=1.0\textwidth]{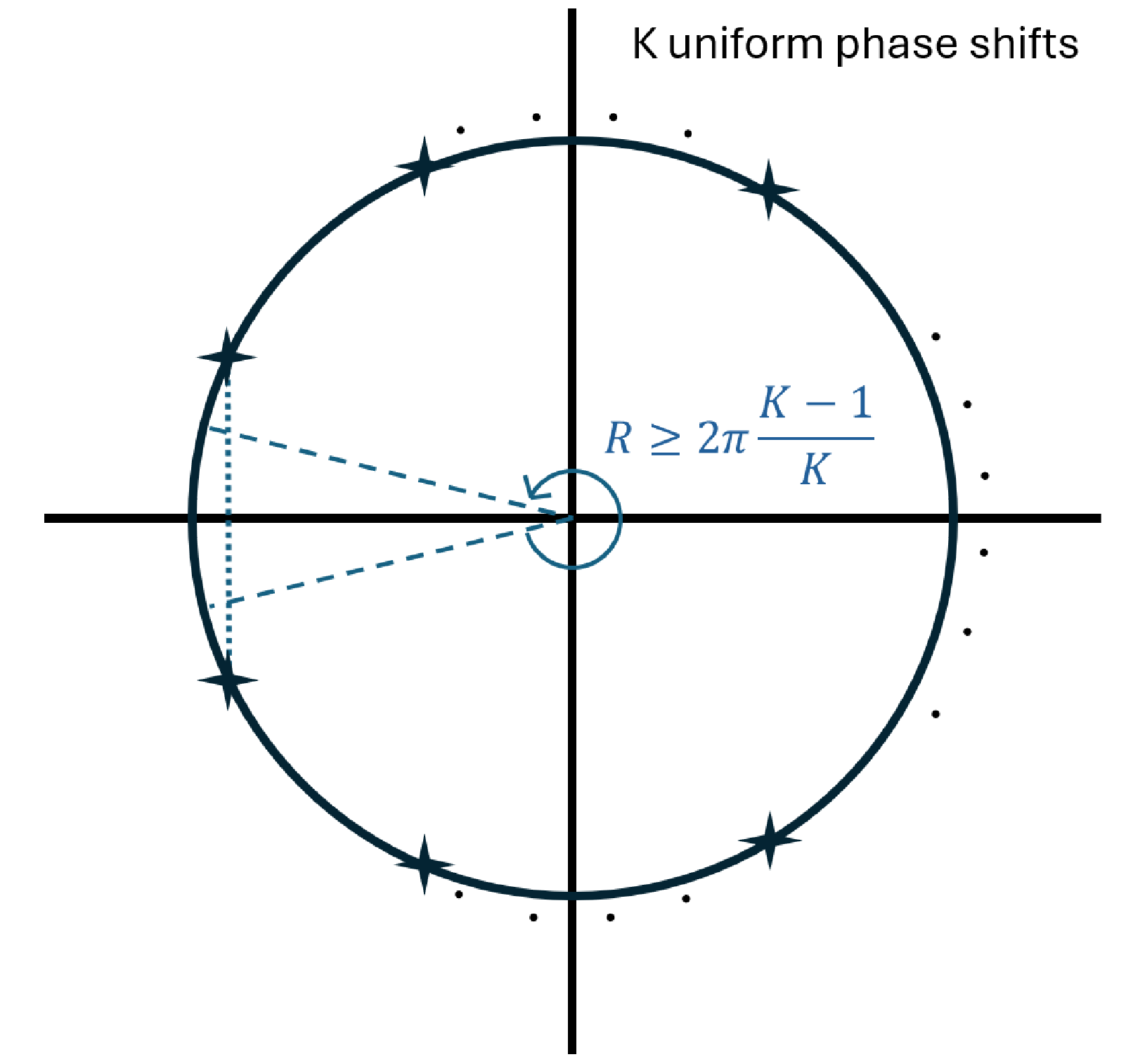}
\caption{Placement of uniformly distributed discrete phase placement for a sufficient phase range.}
 \label{fig:DPS_uniform}
\end{minipage}%
\hspace{0.03\textwidth}
\begin{minipage}{0.48\textwidth}
\centering
\includegraphics[width=1.0\textwidth]{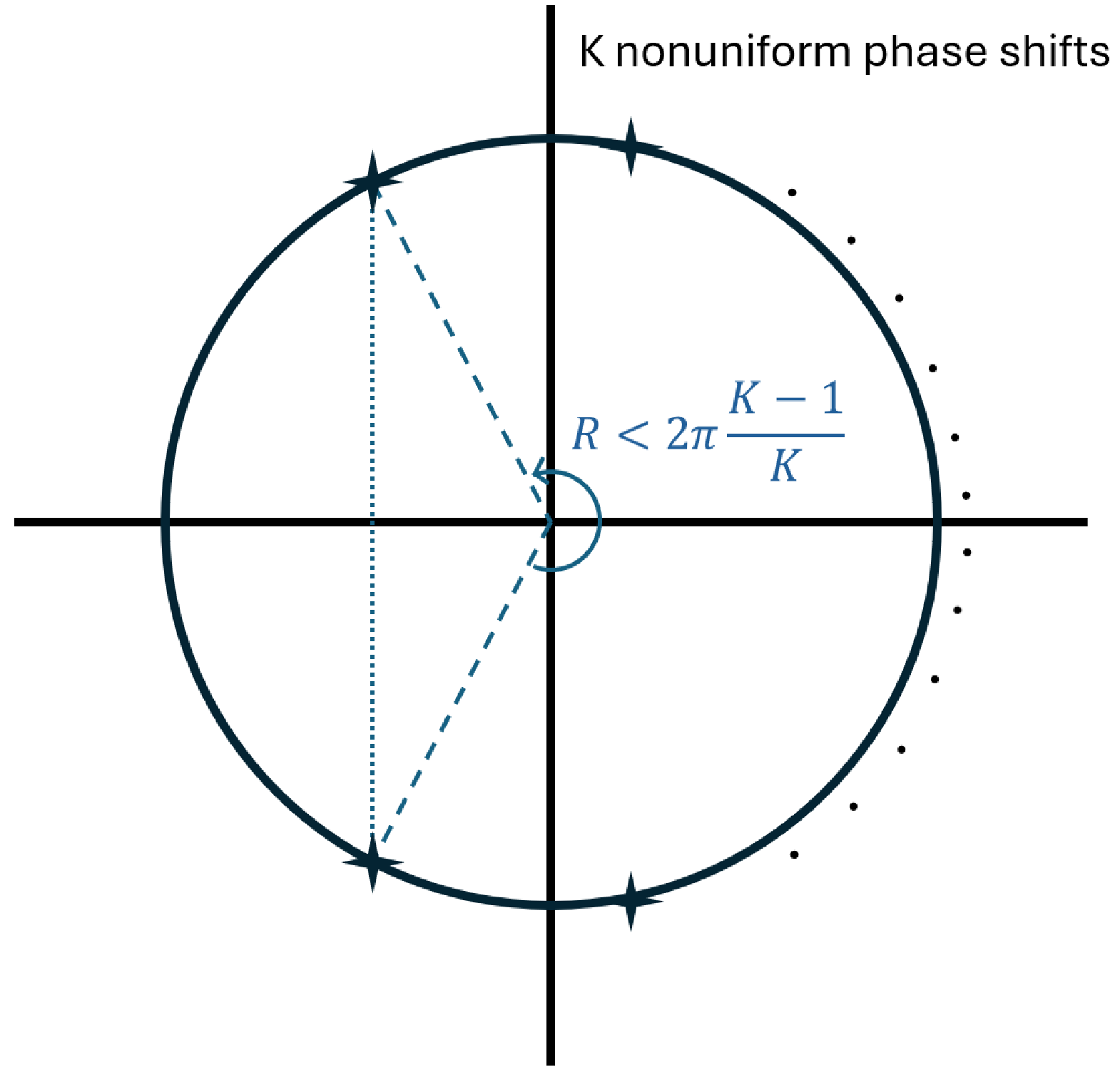}
\caption{Placement of nonuniformly distributed discrete phase placement for a limited phase range.}
 \label{fig:DPS_nonuniform}
\end{minipage}
\end{figure*}
\else
\begin{figure*}[!t]
\centering
\begin{minipage}{0.48\textwidth}
\centering
\includegraphics[width=0.70\textwidth]{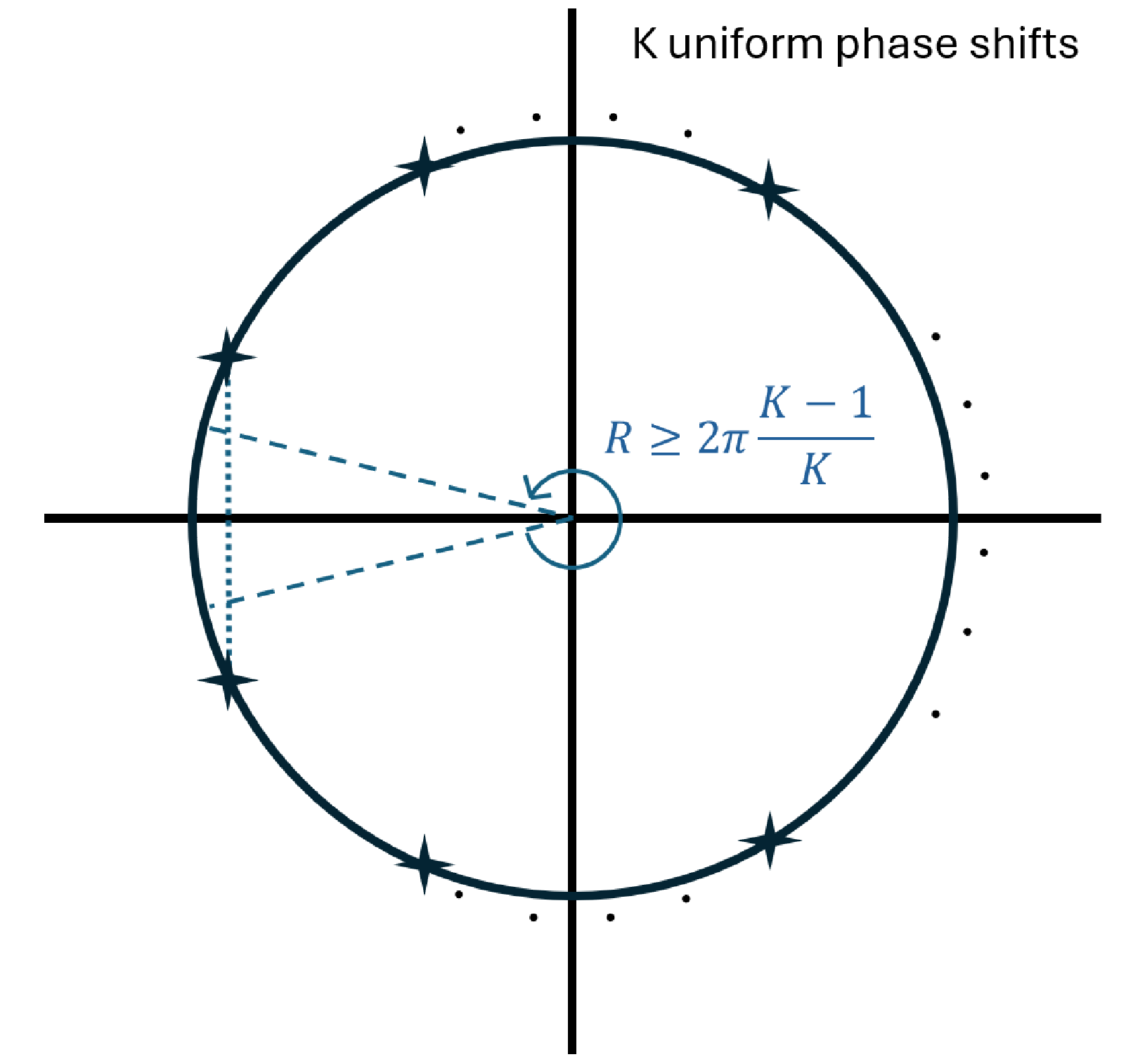}
\caption{Placement of uniformly distributed discrete phase placement for a sufficient phase range.}
 \label{fig:DPS_uniform}
\end{minipage}%
\hspace{0.03\textwidth}
\begin{minipage}{0.48\textwidth}
\centering
\includegraphics[width=0.70\textwidth]{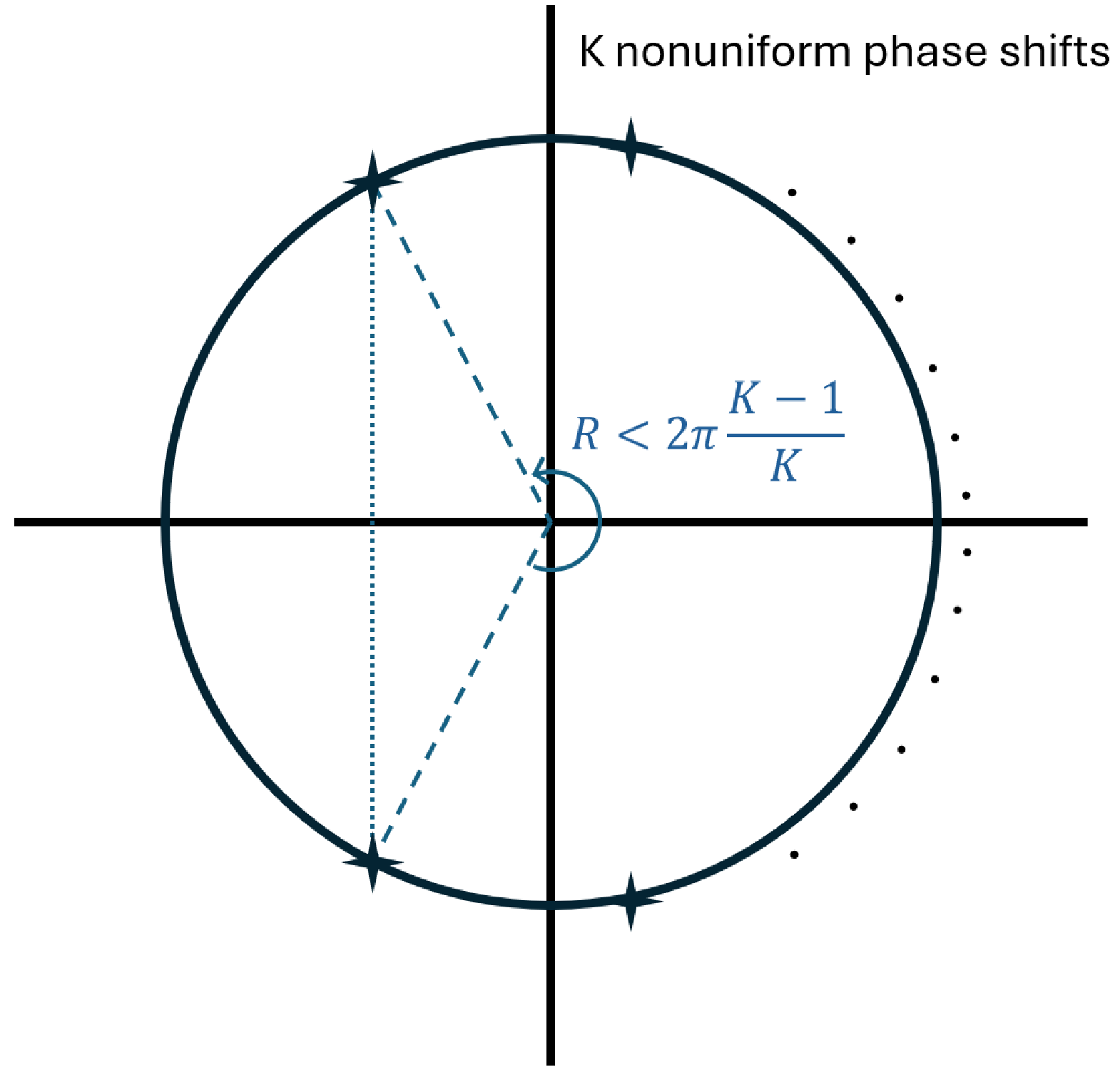}
\caption{Placement of nonuniformly distributed discrete phase placement for a limited phase range.}
 \label{fig:DPS_nonuniform}
\end{minipage}
\end{figure*}
\fi

When it comes to equally separated discrete phases over the RIS phase range, there are two possibilities: the discrete phases may or may not be placed uniformly over the unit circle.\footnote{Note that the terms uniform and nonuniform depend on the range over which they are defined. In this paper, we use the term nonuniform to mean the distribution over the full phase range $[-\pi, \pi)$ is nonuniform, or not equally separated.} 
The condition that determines whether the equally separated discrete phases will be uniform or not depends on the RIS phase range being sufficient or limited, i.e., $R \geq 2\pi\frac{K-1}{K}$ and $R<2\pi\frac{K-1}{K}$, respectively. Therefore, the discrete phase shift set is determined as follows:
\ifCLASSOPTIONonecolumn
\begin{equation}\label{eq:PHIK_PDA}
\Phi_K =
\left\{
\begin{array}{ll}
\{-\frac{R}{2}, \frac{R}{K-1}-\frac{R}{2}, \dots, (K-2)\frac{R}{K-1}-\frac{R}{2}, \frac{R}{2}\} & {\rm for\ } R<2\pi\frac{K-1}{K},\\
\{0, \omega', \dots, (K-1)\omega'\} - \frac{(K-1)\omega'}{2} & {\rm for\ } R \geq 2\pi\frac{K-1}{K},
\end{array}
\right.
\end{equation}
\else
\begin{equation}\label{eq:PHIK_PDA}
\Phi_K =
\left\{
\begin{array}{llll}
\{-\frac{R}{2}, \frac{R}{K-1}-\frac{R}{2}, \dots, (K-2)\frac{R}{K-1}-\frac{R}{2}, \frac{R}{2}\}\\
\hspace{12em}\text{for } R<2\pi\frac{K-1}{K},\\
\{0, \omega', \dots, (K-1)\omega'\} - \frac{(K-1)\omega'}{2}\\
\hspace{12em}\text{for }  R \geq 2\pi\frac{K-1}{K},
\end{array}
\right.
\end{equation}
\fi
where $\omega' = \frac{2\pi}{K}$. Fig.~\ref{fig:DPS_uniform} and Fig.~\ref{fig:DPS_nonuniform} represent the second line and first line of (\ref{eq:PHIK_PDA}), respectively. In Fig.~\ref{fig:DPS_uniform} all arc lengths are the same and equal to $2\pi/K$. Whereas, in Fig.~\ref{fig:DPS_nonuniform}, the arc lengths within the range $R$ are the same and are equal to $R/(K-1)$. In (\ref{eq:PHIK_PDA}), subtracting $\frac{(K-1)\omega'}{2}$ ensures that the uniform discrete phase shifts are symmetric, just like the assumption with the nonuniform discrete phase shifts for compliance. In \cite[Section~VI]{PLA24} it was shown that the placement of $\phi_k$s as in Fig.~\ref{fig:DPS_uniform} and Fig.~\ref{fig:DPS_nonuniform} corresponds to an optimal placement when $\beta^r(\phi_k)=1,$ $k=1, \ldots, K$, for $R \geq 2\pi\frac{K-1}{K}$ and $R<2\pi\frac{K-1}{K}$, respectively.
Therefore, without loss of generality, we determine our discrete phase shifts in the range of $[-\pi, \pi)$ as $-\pi \leq \phi_1 < \cdots < \phi_K < \pi$.
For the performance analysis in this paper, discrete phase shifts will be placed uniformly as shown in Fig.~\ref{fig:DPS_uniform}, if $R$ is large enough, i.e., $R \geq 2\pi\frac{K-1}{K}$. Otherwise, we will use the approach in Fig. \ref{fig:DPS_nonuniform}, i.e., $-\pi \leq \phi_1 < \cdots < \phi_K = \phi_1+R < \pi$ when $R<2\pi\frac{K-1}{K}$.

We remark that, while we provide the possible selection strategies for the discrete phases, these approaches are not required for our algorithms to work. 
The contribution of these assumptions is to enable us to quantify the RIS performance for settings that are suitable for practical scenarios. Without loss of generality, the optimal and suboptimal algorithms proposed in this paper will work with any arbitrary discrete phases to solve the general $K$-ary QP problem.

\subsection{Phase Dependent Amplitude Constraint and A Practical Model}

\begin{figure}[!t]
    \centering
    \includegraphics[width=0.25\textwidth]{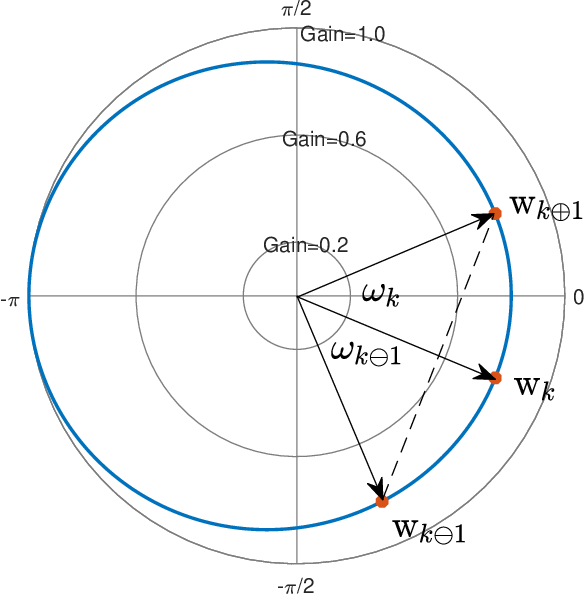}
    \caption{An arbitrary PDA curve on the complex plane with three instances from ${\bf W}_K$ for $K \geq 3$.}
    \label{fig:GeneralPDA_Convexity}
\end{figure}

The PDA attenuation is caused by the RIS elements when they reflect signals, where the resulting gain of the RIS element depends on the selected phase shift of the element \cite{AZWY20, CYTP24}.
In general, $\beta^r(\theta_n): [-\pi, \pi] \rightarrow [0,1]$ can be an arbitrary real-valued function that represents the phase-dependent gain of each RIS element. The analysis and algorithmic framework presented in this paper do not rely on a specific parametric form of $\beta^r(\theta_n)$. Rather, we require a structural condition on the resulting set of RIS coefficients ${\bf W}_K$ to achieve global optimality with the proposed linear-time algorithm. Specifically, for $K \geq 3$, we require that any three consecutive coefficients in ${\bf W}_K$
form a convex triplet, when the triplet spans an angle less than $\pi$. In other words, the middle point does not fall under the line connecting the outer points, thereby preserving convexity.
Under this condition, our search algorithm achieves the globally optimal solution in linear time. When the condition is not met, the algorithm can still yield better performance than the proposed quantization-based method, although the latter remains attractive due to its low complexity and strong approximation behavior. 
Fig.~\ref{fig:GeneralPDA_Convexity} illustrates the PDA function $\beta^r(\theta_n): [-\pi, \pi] \rightarrow [0,1]$, where the $\beta^r(\theta_n)$ values are plotted for $\theta_n \in [-\pi, \pi]$, and the RIS coefficients. The PDA curve is sampled at certain locations to have the resulting RIS coefficients set ${\bf W}_K$. Fig.~\ref{fig:GeneralPDA_Convexity} also shows an example of local convexity among a triplet.

A particularly useful case arises when the continuous gain function $\beta^r(\theta_n)$ traces a strictly convex, smooth, closed curve in the complex plane. For such curves, any arc that spans less than $\pi$ radians lies entirely on the outer side of the chord connecting its end points and exhibits a strictly outward-bulging shape. As a result, for any three consecutive RIS coefficients ${\rm w}_{k \ominus 1}$, ${\rm w}_k$, and ${\rm w}_{k \oplus 1}$ sampled in order along the curve, the middle point ${\rm w}_k$ necessarily lies strictly outside the chord joining ${\rm w}_{k \ominus 1}$ and ${\rm w}_{k \oplus 1}$. 
This follows from the properties of strictly convex curves, which ensure that the arc between any two points lies entirely on one side of the corresponding chord \cite{lay2007convex, schneider2013convex}. 
It is worth noting that strict convexity of $\beta^r(\theta_n)$ is not a necessary condition for the proposed method to apply. Even when the underlying curve is not strictly convex, the sampled discrete set ${\bf W}_K$ often satisfies the convexity condition, depending on the placement of the RIS coefficients. In particular, when the discrete phases are uniformly spaced and $K \leq 4$, the convexity of ${\bf W}_K$ is typically preserved. As a result, the proposed linear-time algorithm remains applicable and achieves the global optimum in most practical settings.

\subsubsection{A Practical PDA Model for RIS}

To analyze the discrete beamforming performance of an RIS in a practical context, we use the mathematical relation for an RIS model provided in \cite{AZWY20} to generate performance results, where the same model \cite{OKWG23, ZZRXA21}, or similar ones \cite{LCLLLW21,CLLL20}, are used in published works on RIS.
According to \cite{AZWY20}, the relation between the phase-dependent gain $\beta^r(\theta_n)$ and the phase shift $\theta_n$ for the $n$-th RIS element is
\begin{equation}\label{eq:PDAmodel}
\beta^r(\theta_n) = (1 - \beta^r_{min}) \left(\frac{\sin(\theta_n - \phi^r) + 1}{2}\right)^{\alpha^r} + \beta^r_{min}
\end{equation} 
where $\beta^r_{min}$ is the minimum amplitude corresponding to the maximum attenuation and therefore quantifies the level of attenuation, e.g., the lower the $\beta^r_{min}$ the higher the attenuation. Variables $\beta^r_{min}$, $\phi^r$, and $\alpha^r$ are practical hardware-related parameters that depend on the implementation of the RIS and are independent of the channel. In (\ref{eq:PDAmodel}), $\phi^r \geq 0$ determines the phase shift for which the maximum attenuation occurs. It is defined as the difference between $-\pi/2$ and $\theta_n$ such that $\beta^r(\theta_n = \phi^r - \pi/2) = \beta^r_{min}$. The parameter $\alpha^r \geq 0$ controls the steepness of the PDA curve, governing the transition from the maximum gain at $\phi^r + \pi/2$ to the minimum gain at $\phi^r - \pi/2$. 
The selection of $\phi^r$ essentially rotates the whole PDA curve on the complex plane. Choosing $\pi/2$ makes the PDA curves symmetric around the real axis, similar to discrete phase shift selections. We use $\phi^r = \frac{\pi}{2}$ in the remainder of this paper. This assumption helps with the flow and, as will be discussed in the sequel, the calculation of the approximation ratio for the RIS. 
Among the PDA parameters, $\beta^r_{min}$ plays the most significant role, therefore we focus our analysis on this parameter. With this, it is readily shown in \cite{AZWY20} that the selection of $\alpha^r$ has a marginal impact on the overall performance compared to $\beta^r_{min}$. Therefore, we fix $\alpha^r = 1.6$ to get the results in this paper, similar to \cite{AZWY20}.

\subsubsection{Discussion on Convexity of the Practical RIS Model}

The shape of the PDA curve is primarily determined by two parameters: $\beta^r_{min}$ and $\alpha^r$, while $\phi^r$ induces a rotation of the entire curve. 
We closely examined the convexity of the PDA curve.
Under the PDA constraint, a convex PDA profile is generally maintained, except in extreme cases with significant attenuation, e.g., when $\beta^r_{min} < 0.4$ and $\alpha^r < 1.2$. Similarly, the convexity is also preserved for $\beta^r_{min} < 0.4$, provided that $\alpha^r$ does not become excessively large, e.g., $\alpha^r \leq 2$.

To ensure convex RIS coefficients across a wide range of scenarios, we point to the values $\beta^r_{\min} \in [0.4, 1]$ and $\alpha^r \in [1.4, 2]$. This range is both practical and effective for preserving convexity, and the choice of $\alpha^r$ offers flexibility due to its marginal impact on system performance \cite{AZWY20}. Finally, we remark that, prior works addressing the problem in (\ref{eqn:problem1_pda}) with uniform and nonuniform discrete phase shifts, i.e., \cite{PA24} and \cite{PLA24}, respectively, exhibited global convexity in the RIS coefficients due to the assumption of constant gain at the RIS elements, i.e., having a circular PDA curve. 


In the following sections, we assume convex RIS coefficients in the remainder of this paper, based on the analysis in this section. Having the problem and the constraints established, we will next provide optimal and suboptimal algorithms to solve the received power maximization problem in (\ref{eqn:problem1_pda}). 
\section{Optimal Solution with Discrete Phase Shifts}\label{ch:globalOptimum}
In this section, we provide the necessary and sufficient condition to achieve the global optimum solution. Based on the convexity discussions on the RIS coefficients ${\bf W}_K$ in Section~\ref{sec:PDAandPHASERANGE_MODEL}, we will employ the necessary and sufficient conditions to get the global optimum in linear time. Note that we want to maximize $| h_0 + \sum_{n=1}^N h_n \beta^r(\theta_n) e^{j\theta_n} |^2$ where $\theta_n \in \Phi_K$ and $h_n = \beta_ne^{j\alpha_n}$ for $n = 0, 1, \ldots, N$, $\beta_n \geq 0$, and $\alpha_n \in [-\pi, \pi)$. Let $\theta^*_n$ for $n=1, \ldots, N$ be the discrete phase shift selections that give the global optimum. Define $g$ as
\begin{equation}\label{eqn:gstar}
g = h_0 + \sum_{n=1}^N h_n \beta^r(\theta^*_n)e^{j\theta^*_n}.
\end{equation} 
Let $\mu=g/|g|$ such that $|g| = g e^{-j \angle\mu}$. Similar to the condition in \cite{PLA24}, we define the following lemma.

{\em Lemma~1:\/} For an optimal solution $(\theta_1^*, \theta_2^*, \ldots,
\theta_n^*)$ for the received power maximization problem given in (\ref{eqn:problem1_pda}), it is necessary and sufficient that each $\theta_n^*$ satisfy
\begin{equation}
\theta_n^* = \arg \max_{\theta_n\in \Phi_K} \beta^r(\theta_n) \cos(\theta_n + \alpha_n -\phase{\mu})
\label{eqn:lemma1_pda}
\end{equation}
for an arbitrary $\Phi_K$.

{\em Proof:\/} We can rewrite $|g| = g e^{-j \angle\mu}$ as
\begin{align}
|g| =& \ \beta_0 e^{j(\alpha_0-\phase{\mu})} + \sum_{n=1}^N \beta_n \beta^r(\theta^*_n) e^{j(\alpha_n+\theta_n^*-\phase{\mu})} \\
 = & \ \beta_0 \cos(\alpha_0 - \phase{\mu}) + j \beta_0 \sin (\alpha_0-\phase{\mu}) \nonumber\\
& + \sum_{n=1}^N \beta_n \beta^r(\theta^*_n) \cos(\theta_n^* + \alpha_n - \phase{\mu}) \nonumber\\
& + j \sum_{n=1}^N \beta_n \beta^r(\theta^*_n) \sin(\theta_n^* + \alpha_n - \phase{\mu}).
\label{eqn:absg_pda}
\end{align}
Because $|g|$ is real-valued, the second and fourth terms in (\ref{eqn:absg_pda}) sum to zero, and
\begin{equation}\label{eq:resulting_g}
|g| = \beta_0 \cos(\alpha_0 - \phase{\mu}) + \sum_{n=1}^N \beta_n \beta^r(\theta^*_n) \cos(\theta_n^* + \alpha_n - \phase{\mu}),
\end{equation}
from which (\ref{eqn:lemma1_pda}) follows as a necessary and sufficient condition for the lemma to hold, since $\beta_n \geq 0$ for $n = 1, \ldots, N$.
\hfill$\blacksquare$

We remark that, unlike our previous works on the received power maximization problem \cite{PA24, PLA24}, maximizing $|g|$ in (\ref{eq:resulting_g}) is actually difficult with continuous phases. With discrete phases, one could at least attempt an exhaustive search to find the global optimum. 
Whereas, with continuous phases, finding the global optimum requires taking the partial derivatives of (\ref{eq:resulting_g}) with respect to each $\theta_n$. However, $\phase{\mu}$ is also a function of $\theta_n$ and this complicates the derivation significantly. 
Exhaustive search would be prohibitive as it would take $\mathcal{O}(K^N)$ steps where RISs are considered to have hundreds or even thousands of elements $N$. Yet, we will come up with a linear-time algorithm in the remainder of this section, enabling the received power maximization with discrete phase shifts. 

\begin{figure}[!t]
	\centering	
    \includegraphics[width=0.40\textwidth]{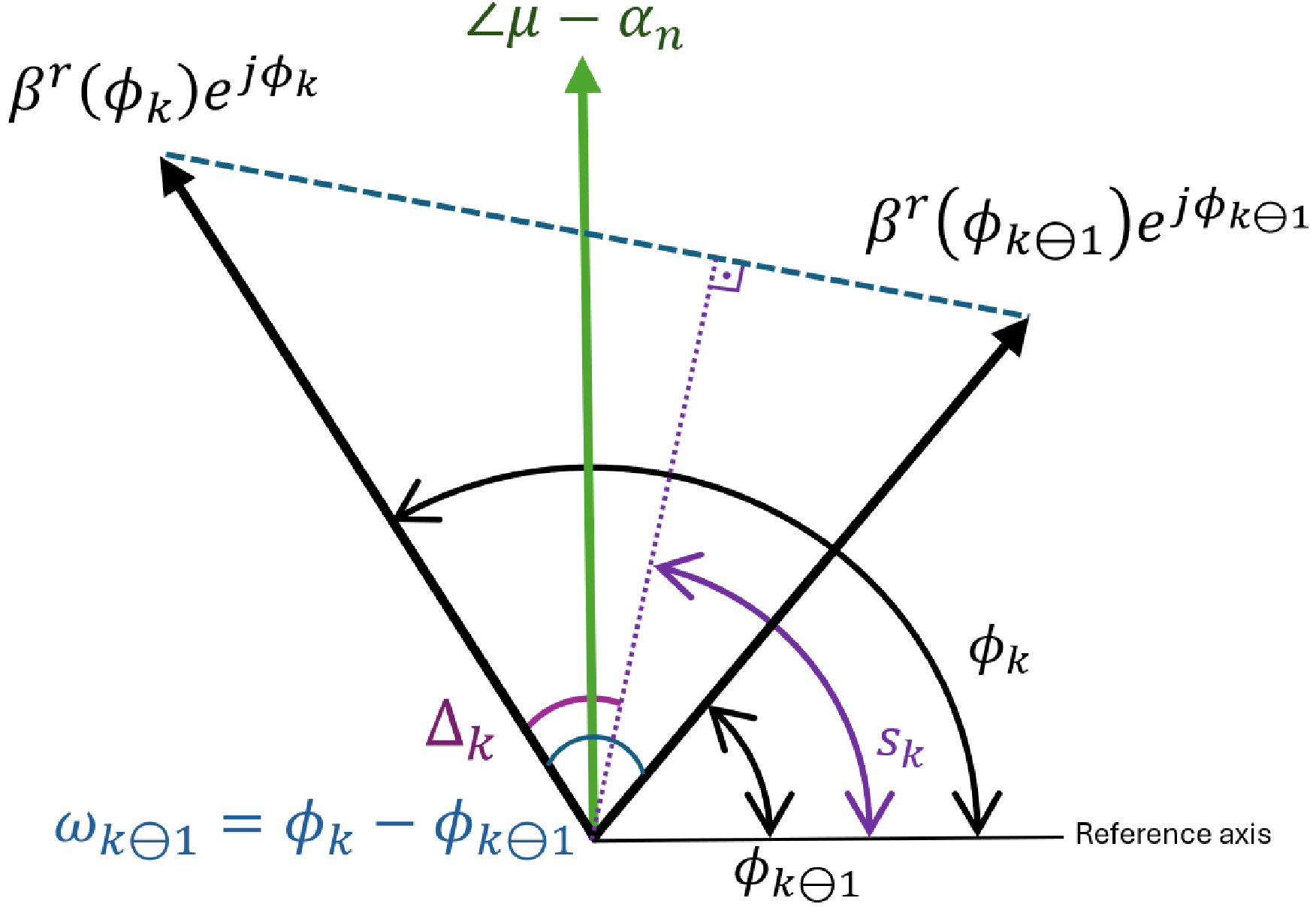}
	\caption{A simple case of the new decision boundary. Decision boundary: Purple - at angle $s_k$ - for $\phi_{k\ominus1} \rightarrow \phi_k$.}
	\label{fig:decisionBoundary_simple}
\end{figure}

Note that, in order to use the necessary and sufficient condition to achieve the global optimum with discrete phases, we need to know $\phase{\mu}$, which comes from the optimal discrete phases that we do not know yet. Unless we have another condition, there are infinitely many possible $\phase{\mu}$ values over the range $[-\pi, \pi)$. To make this into a finite set of possibilities, we start by observing the objective term in (\ref{eqn:lemma1_pda}). The angle $(\theta_n + \alpha_n - \phase{\mu})$ inside the cosine corresponds to the angle between the rotated channel vector and optimum $|g|$. Therefore, the overall value $\beta^r(\theta_n) \cos(\theta_n + \alpha_n - \phase{\mu})$ corresponds to the projection of the rotated and scaled channel vector on the resulting sum. In other words, it is necessary and sufficient that for each RIS element, we need to maximize the length of the projection of the rotated and scaled channel vector. For better understanding, we use the simplified vector diagram in Fig.~\ref{fig:decisionBoundary_simple} to illustrate a decision boundary. 
For the $n$-th RIS element, we use $(\phase{\mu} - \alpha_n)$ to represent the optimal direction and omit the channel gain $\beta_n$ as it is independent from the selected phase shifts. The boundary between the two options will be the dashed red line that is drawn perpendicular from the origin to the line connecting the two RIS coefficients. If 
$(\phase{\mu} - \alpha_n)$
is before this line, $\theta_n = \phi_{k\ominus1}$ is chosen. Similarly, if 
$(\phase{\mu} - \alpha_n)$
is after this line, $\theta_n = \phi_{k}$ is chosen.
To come up with a finite set of options for $\phase{\mu}$, we need to find the boundary between each adjacent RIS coefficient, which amounts to $K$ boundaries per element. Let $s_k$ for $k=1, \ldots, K$ denote the angle of these boundaries and let $\Delta_k$ be the angle from $s_k$ to $\phi_k$, then we have
\begin{equation*}
    \beta^r(\phi_k)\cos(\Delta_k) = \beta^r(\phi_{k\ominus1})\cos(\omega_{k\ominus1} - \Delta_k),
\end{equation*}
meaning
\begin{equation*}
    \tan(\Delta_k) = \frac{\beta^r(\phi_k)-\beta^r(\phi_{k\ominus1})\cos(\omega_{k\ominus1})}{\beta^r(\phi_{k\ominus1})\sin(\omega_{k\ominus1})},
\end{equation*}
and
\begin{equation*}
    \Delta_k = \arctan\left(\frac{\beta^r(\phi_k)-\beta^r(\phi_{k\ominus1})\cos(\omega_{k\ominus1})}{\beta^r(\phi_{k\ominus1})\sin(\omega_{k\ominus1})}\right),
\end{equation*}
and, as a result,
\begin{equation}\label{eq:sk_boundary}
    s_k = \phi_k - \arctan\left(\frac{\beta^r(\phi_k)-\beta^r(\phi_{k\ominus1})\cos(\omega_{k\ominus1})}{\beta^r(\phi_{k\ominus1})\sin(\omega_{k\ominus1})}\right).
\end{equation}
Therefore, given the boundaries at $s_k$ for the $n$-th RIS element, we define the following sequence of complex numbers for each $n=1, 2, \ldots, N$ to represent each and every boundary on the unit circle
\begin{equation} \label{eq:snk_pda}
    s_{nk} = e^{j\left(\alpha_n + \phi_k - \arctan\left(\frac{\beta^r(\phi_k)-\beta^r(\phi_{k\ominus1})\cos(\omega_{k\ominus1})}{\beta^r(\phi_{k\ominus1})\sin(\omega_{k\ominus1})}\right)\right)}
\end{equation} for $k = 1, 2, \ldots, K$.

To show the simplified version of the boundaries when there is no gain attenuation, let $\beta^r_{min} = 1$ to have unit gain among all the RIS elements, that is, $\beta^r(\phi_{k\ominus1}) = \beta^r(\phi_k)$. From the half-angle identity, this would result in $\arctan\left(\frac{1-\cos(\omega_{k\ominus1})}{\sin(\omega_{k\ominus1})}\right) = \frac{\omega_{k\ominus1}}{2}$ and $s_{nk} = \exp(j(\alpha_n + \phi_k - \frac{\omega_{k\ominus1}}{2}))$, which turn out to be the same decision boundaries as in \cite{PLA24}. This shows that the decision boundaries and the respective solutions in \cite{PLA24} with the ideal gain assumption are a special case of the non-ideal gains considered in this paper.

\begin{figure}[!t]
	\centering
    \includegraphics[width=0.50\textwidth]{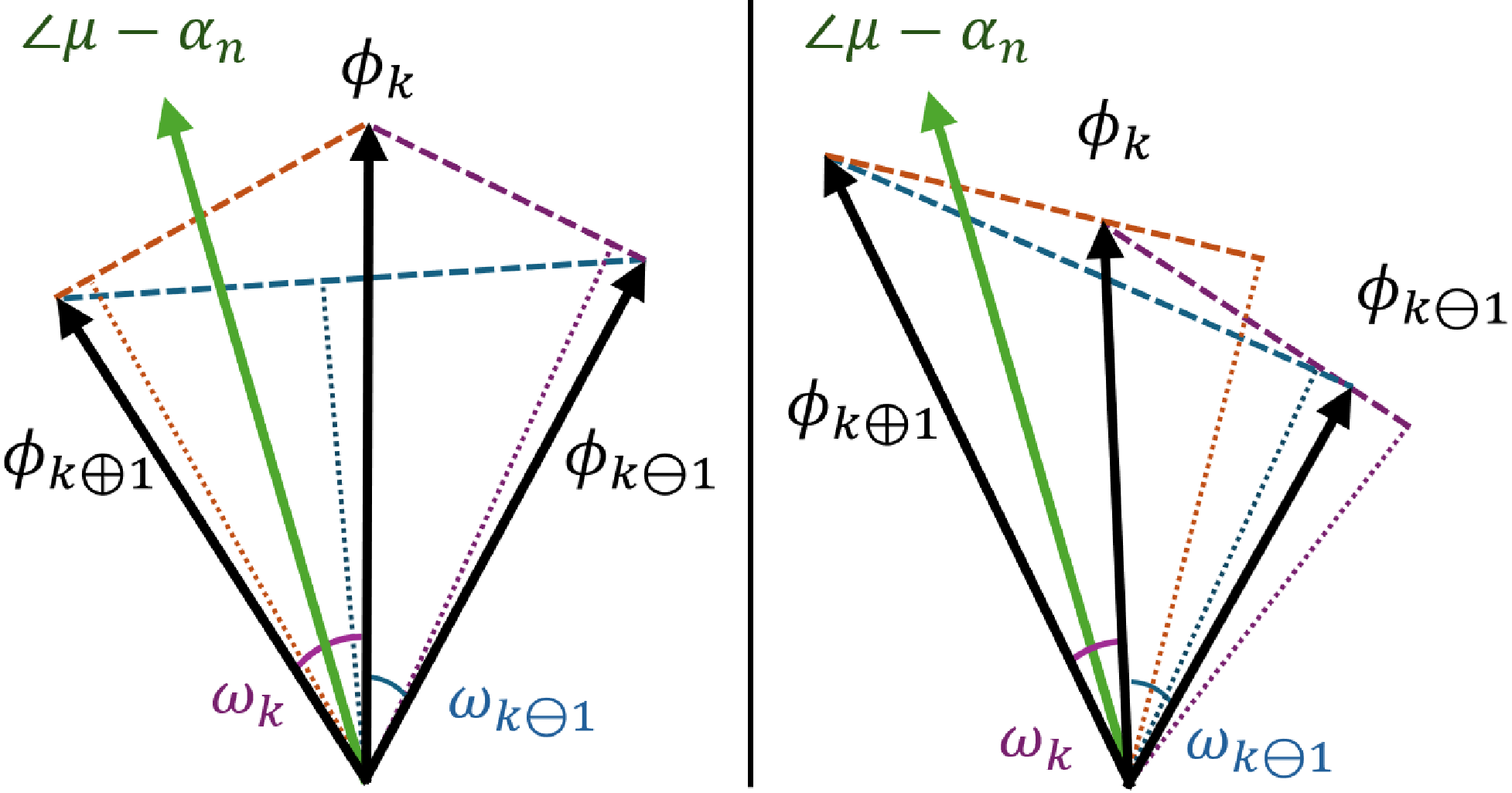}
	\caption{An illustration for the convexity for the RIS coefficients. Decision boundaries: Purple - at angle $s_k$ - for $\phi_{k\ominus1} \rightarrow \phi_k$; Orange - at angle $s_{k\oplus1}$ - for $\phi_k \rightarrow \phi_{k\oplus1}$; Blue - for $\phi_{k\ominus1} \rightarrow \phi_{k\oplus1}$.}
	\label{fig:convexCases}
\end{figure}

We provide an additional sufficiency condition on $\theta^*_n$ using $s_{nk}$ in (\ref{eq:snk_pda}), which will help us reduce the search space of $\mu$, and come up with a linear time algorithm to achieve the global optimum solution, similar to \cite{PA24, PLA24}. 
For this purpose, we assume for $K \geq 3$ that 
any three consecutive coefficients in ${\bf W}_K$ 
follow a convex curve as shown in Fig.~\ref{fig:convexCases}, i.e., with respect to the origin, the coefficient in the middle is above the line connecting the adjacent coefficients. 
We note that this is actually a realistic assumption rather than an optimistic one, as discussed in Section~\ref{sec:PDAandPHASERANGE_MODEL}.
Let us define ${\rm arc}(a:b)$, for any two points $a$ and $b$ on the unit circle $C$,  to be the unit circular arc with $a$ as the initial end and $b$ as the terminal end in the counterclockwise direction, with the two endpoints $a$ and $b$ being excluded.

{\em Proposition 1:\/} A sufficient condition for $\theta_n^*=\phi_k$, independent of $K$, is
\begin{equation}
\mu \in {\rm arc} (s_{nk}:s_{n,k\oplus1}).
\label{eqn:prop1}
\end{equation}

\begin{figure}[!t]
	\centering
    \includegraphics[width=0.48\textwidth]{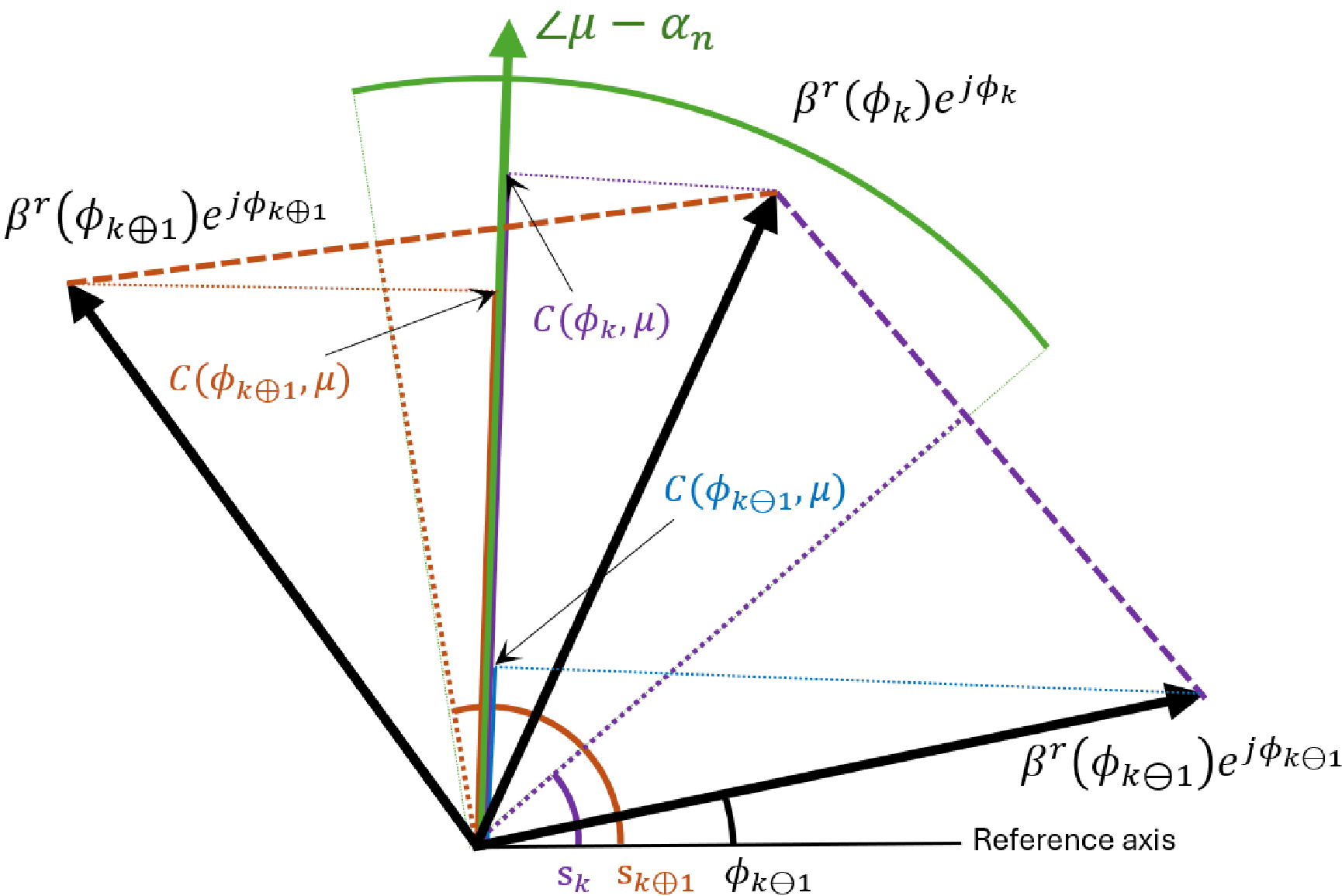}
	\caption{An illustration for the optimality of $\theta_n^*=\phi_k$ given $\mu \in {\rm arc} (s_{nk}:s_{n,k\oplus1})$. Decision boundaries: Purple - at angle $s_k$ - for $\phi_{k\ominus1} \rightarrow \phi_k$; Orange - at angle $s_{k\oplus1}$ - for $\phi_k \rightarrow \phi_{k\oplus1}$.}
	\label{fig:prop1_pda}
\end{figure}

{\em Proof:\/} Assume $\mu$ satisfies (\ref{eqn:prop1}). Then,
\begin{equation}\label{eq:muRange1}
    \phase\mu \in \left(\alpha_n + \phi_k - \Delta_k, \alpha_n + \phi_{k\oplus1} - \Delta_{k\oplus1}\right).
\end{equation}
where $\Delta_k = \arctan(\frac{\beta^r(\phi_k)-\beta^r(\phi_{k\ominus1})\cos(\omega_{k\ominus1})}{\beta^r(\phi_{k\ominus1})\sin(\omega_{k\ominus1})})$. By taking the negation of (\ref{eq:muRange1}) and adding $\theta_n$ and $\alpha_n$, we get
\begin{equation}
    \theta_n+\alpha_n-\phase\mu \in \left(\theta_n + \Delta_{k\oplus1}-\phi_{k\oplus1}, \theta_n + \Delta_k-\phi_k\right).
\end{equation}
Now, consider each case when $\theta_n \in \{\dots, \phi_{k\ominus1}, \phi_k, \phi_{k\oplus1}, \dots\}$ and compute the projection of the rotated and scaled channel through the $n$-th RIS element onto the optimal direction $\phase{\mu}$. 
We define a contribution term to represent this projection as
\begin{equation}\label{eq:contribution_prop1}
    C(\theta_n,\mu) \triangleq \beta^r(\theta_n)\cos(\theta_n - (\phase{\mu} - \alpha_n)).
\end{equation}
Fig.~\ref{fig:prop1_pda} provides a visual illustration of the aforementioned projection, excluding the channel gain term $\beta_n$, as it scales all phase shift selections equally. Note that, $C(\theta_n,\mu)$ correspond to the objective term in {\em Lemma~1}. 
Consider the two edge cases where $(\phase{\mu} - \alpha_n)$ is close to either $s_k$ or $s_{k\oplus1}$. In these scenarios, the definition of $s_k$ in (\ref{eq:sk_boundary}) guarantees that $C(\phi_{k\ominus1},\mu)<C(\phi_k,\mu)$ and $C(\phi_{k\oplus1},\mu)<C(\phi_k,\mu)$, respectively. 
Therefore, given $(\phase{\mu} - \alpha_n) \in (s_k, s_{k\oplus1})$, i.e., $\mu \in {\rm arc} (s_{nk}:s_{n,k\oplus1})$, any phase shift selection other than $\theta_n = \phi_k$ will evidently result in a lower projection value. 
Therefore, the contribution $C(\theta_n = \phi_k,\mu)$ is the maximum among all possible choices, ensuring that (\ref{eqn:prop1}) satisfies the necessary and sufficient condition, thus completing the proof.
\hfill$\blacksquare$

Finally, to operate with Proposition~1, we will eliminate duplicates among $s_{nk}$ and sort to get $e^{j\lambda_l}$ such that $0\le \lambda_1 < \lambda_2 < \cdots < \lambda_L < 2\pi$ as in \cite{PA24} and \cite{PLA24}. To achieve the optimum solution in linear time, instead of assigning each $\theta_n$ with {\em Lemma~1} for each candidate $\mu$, we will utilize {\em Proposition~1} to develop a search algorithm with elementwise updates. For this purpose, we need to track $\mu$ switching from one arc to another, i.e.,
\begin{equation}
\mu\in {\rm arc}{(e^{j\lambda_l}:e^{j\lambda_{l+1}})} \rightarrow \mu\in {\rm arc}(e^{j\lambda_{l+1}}:e^{j\lambda_{l+2}}),
\label{eq:muSwitch}
\end{equation}
and update the respective $\theta_n$ as given by the sufficiency condition in {\em Proposition~1}. With this, the elementwise update rule can be defined as
\begin{equation}\label{eq:updateRule}
    {\cal N} (\lambda_l) = \{ \{n',k'\} | \phase{s_{n'k'}} = \lambda_l \}.
\end{equation}

The optimum algorithm starts by intializing a candidate $\phase{\mu}$, say $\phase{\mu} = 0$. At first, the discrete phases $\theta_n$ for $n=1, \dots, N$ are initialized with {\em Lemma~1}. Then $\mu$ traverses the unit circle in the counterclockwise direction. As $\mu$ moves along the unit circle and jumps over a boundary $e^{j\lambda_{l+1}}$, $\theta_n'$ will be updated for every $\{n',k'\} \in {\cal N} (\lambda_{l+1})$ according to the update rule in (\ref{eq:updateRule}) as
\begin{equation}
\theta_{n'} \rightarrow \phi_{k'}, \quad \{n',k'\} \in {\cal N}(\lambda_{l+1}).
\label{eqn:eqn41}
\end{equation}
Therefore, it is sufficient to consider $L \leq NK$ steps to find the global optimum, where only one or a few phase shifts are updated. This procedure is specified under Algorithm~1, which is a generalized version of Algorithm~1 from \cite{PLA24} to work with the PDA constraint, or in general for arbitrary RIS coefficients ${\bf W}_K$ that are locally convex.
\begin{algorithm}[!t]
\caption{Generalized Algorithm~1 \cite{PLA24} for Phase-Dependent Amplitude}\label{alg:PDA_optimum_convex}
\begin{algorithmic}[1]
\State {\bf Initialization:} Compute 
$s_{nk}$ and ${\cal N} (\lambda_l)$ as in equations (\ref{eq:snk_pda}) and (\ref{eq:updateRule}), respectively.
\State 
Set $\phase{\mu} = 0$. For $n=1,2,\ldots,N$, calculate and store
\[
\theta_n = \arg\max_{\theta_n\in\Phi_K} \beta^r(\theta_n) \cos(\theta_n + \alpha_n -\phase{\mu}).
\]
\State Set $g_0 = h_0 + \sum_{n=1}^N h_n \beta^r(\theta_n) e^{j\theta_n}$, ${\tt absgmax} = |g_0|$.
\For{$l = 1, 2, \ldots, L-1$}
\State For each double $\{n',k'\} \in {\cal N}(\lambda_l)$, let $\theta_{n'} = \phi_{k'}$.
\State Let
\small
\[
g_l = g_{l-1} + \hspace{-1.75em} \sum_{\{n',k'\}\in{\cal N}(\lambda_l)} \hspace{-1.5em}h_{n'} \big(\beta^r(\theta_n) e^{j\theta_n} - \beta^r(\phi_{k'\ominus1}) e^{j (\phi_{k'\ominus1}) } \big)
\]
\normalsize
\If{$|g_l| > {\tt absgmax}$}
\State Let ${\tt absgmax} = |g_l|$
\State Store $\theta_n$ for $n=1,2,\ldots,N$
\EndIf
\EndFor
\State Read out $\theta_n^*$ as the stored $\theta_n$, $n=1,2,\ldots,N$.
\end{algorithmic}
\end{algorithm}
\begin{figure}[!t]
\centering
\includegraphics[width=0.45\textwidth]{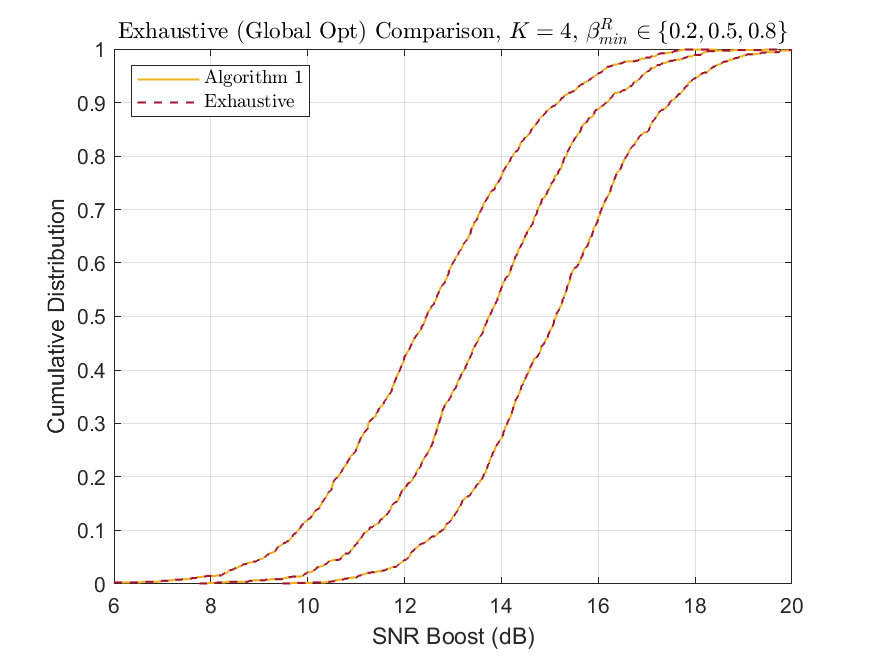}
\caption{CDF plots for SNR Boost with Algorithm~1 and exhaustive search, for $K=4$ and $\beta^r_{min} \in \{0.2, 0.5, 0.8\}$, $N=10$.}
\label{fig:pda_exha_cdf_k4}
\end{figure}

We present the cumulative distribution function (CDF) results for SNR Boost \cite{b1} in Fig.~\ref{fig:pda_exha_cdf_k4} for uniform discrete phases with $K=4$ and $\beta^r_{min} \in \{0.2, 0.5, 0.8\}$. Considering these low, medium, and high values for $\beta^r_{min}$, we compare Algorithm~1 with the exhaustive search results as a numerical validation of the optimality for $N=10$. The CDF plots are generated with $1,000$ channel realizations where both Algorithm~1 and exhaustive search ran over the same realization in each step. It is clear that the results of Algorithm~1 perfectly align with the exhaustive search results, which are optimum.


%
\section{Quantization Approaches}\label{ch:quantization}

In this section, we propose two intuitive quantization algorithms. The first algorithm will be the amplitude-introduced polar quantization (APQ) algorithm that is similar to the straightforward quantization approach. APQ will help us derive the approximation ratio to better understand the capability of the RIS under phase-dependent amplitude constraints and a limited phase range, compared to the ideal scenario with continuous phases discussed in Section~\ref{sec:PDAandPHASERANGE_MODEL}. 
Second, we will define the extended version APQ, namely the extended amplitude-introduced polar quantization (EAPQ) algorithm. Similarly to APQ, EAPQ is also an intuitive algorithm, but unlike the APQ approach, EAPQ will be aware of the amplitude attenuation. In a sense, EAPQ aims to approximate the global optimum solution provided by Algorithm~1.

\subsection{Amplitude-introduced Polar Quantization}

The intuitive quantization algorithm is based on the ideal scenario discussed in Section~\ref{sec:PDAandPHASERANGE_MODEL}, with continuous phase shifts and $\beta^r_{min}=1$, i.e., no amplitude attenuation. The continuous solution for the ideal scenario provides the maximum achievable result. To formally define the ideal solution, similar to \cite{PLA24}, let us first relax $\theta_n$ and rewrite the maximization problem in (\ref{eqn:problem1_pda}) as follows:
\begin{equation}
    \begin{aligned}
        & \underset{\boldsymbol{\theta}^\text{ideal}}{\rm maximize\ } f(\boldsymbol{\theta}^\text{ideal})\\
        & {\rm subject\ to\ } \beta^r(\psi) = 1 ,\ \text{for all} \ \psi \\
        &\,\,\,\,\,\,\,\,\,\,\,\,\,\,\,\,\,\,\,\,\,\,\,\,\,\ \theta_n^{\text{ideal}} \in [-\pi,\pi),\ n=1, 2, \ldots, N,
    \end{aligned}
    \label{eqn:eqn1rx}
\end{equation}
where (\ref{eq:finitial_theta_pda}) is rewritten as 
\begin{equation}
f(\boldsymbol{\theta}^\text{ideal}) = \bigg|\beta_0e^{j\alpha_0}+\sum_{n=1}^N \beta_n \beta^r(\theta_n^{\text{ideal}})
e^{j(\alpha_n + \theta_n^{\text{ideal}})}\bigg|^2.
\end{equation}

Note that, in $f(\boldsymbol{\theta}^\text{ideal})$, all RIS elements will have unit gains and this is because the ideal scenario assumes $\beta^r_{min}=1$ which corresponds to the outermost gray unit circle in Fig.~\ref{fig:GeneralPDA_Convexity}. Therefore, with $\beta^r(\theta_n^{\text{ideal}})=1$ for $n=1,\dots,N$, it is trivial that setting
\begin{equation}
	\theta_n^{\text{ideal}} = \alpha_0 - \alpha_n,\ {\rm for}\  n=1,2,\dots,N
	\label{eq:thetaIdeal}
\end{equation}
aligns all cascaded channels with the direct link $h_0$. In the absence of the direct link, the phase of the direct channel can be assumed zero, i.e., $\alpha_0 = 0$, and the same maximization result can be achieved. With (\ref{eq:thetaIdeal}) and the ideal scenario, RIS will perfectly align all complex vectors in (\ref{eq:finitial_theta_pda}) achieving the maximum possible gain for the given channel, which is $f(\boldsymbol{\theta}^\text{ideal}) = (\sum_{n=0}^{N}\beta_n)^2$. Finally, we remark that for a given $\alpha_0$, $\theta_n^{\text{ideal}}$ for $n=1, \dots, N$ are i.i.d. uniform random variables, i.e., $\theta_n^{\text{ideal}} \sim {\cal U}[-\pi. \pi]$.

To satisfy the discrete phase shift constraint, the APQ algorithm will quantize the ideal solution. Therefore, similar to UPQ in \cite{PA24} and NPQ in \cite{PLA24}, the decision rule for APQ given the ideal solution in (\ref{eq:thetaIdeal}) is defined as
\begin{equation}\label{eq:thetaNPQ}
\theta_n^{\text{APQ}} = 
\left\{
\begin{array}{ll}
\phi_1  & \,\, \text{if}\,\,\,\, \frac{2\pi-\phi_K+\phi_1}{2} \leq \theta_n^{\text{ideal}} < \frac{\phi_1+\phi_2}{2} \\
\phi_2  & \,\, \text{if}\,\,\,\, \quad \quad \,\,\,\, \frac{\phi_1+\phi_2}{2} \leq \theta_n^{\text{ideal}} < \frac{\phi_2+\phi_3}{2} \\
& \vdots \\
\phi_{K-1} & \,\, \text{if}\,\,\,\, \frac{\phi_{K-2}+\phi_{K-1}}{2} \leq \theta_n^{\text{ideal}} < \frac{\phi_{K-1}+\phi_{K}}{2} \\
\phi_{K} & \,\, \text{otherwise}.
\end{array}
\right.
\end{equation}

Note that, from the definition of $\theta_n^{\text{APQ}}$, APQ will select the discrete phase shift that maximizes $\cos(\theta_n + \alpha_n - \alpha_0)$, where $\alpha_0$ replaced $\phase{\mu}$. In a sense, APQ ignores the amplitude attenuation and instead of using $\mu$, which needs to be searched, it adopts $\alpha_0$ and performs the selection. But, even with a coincidence of $e^{-j \angle\mu}$ and $e^{j\alpha_0}$ being in the same arc, APQ may still fail to achieve the global optimum since it ignores the amplitude attenuation in RIS gains.

\subsection{Extended Amplitude-introduced Polar Quantization}

To improve the APQ algorithm that ignores the amplitude attenuation and just focuses on the quantization, we introduce a projection-based approach similar to Algorithm~1, and name this extended algorithm as extended amplitude-introduced polar quantization (EAPQ). Therefore, we consider the projection onto the direct channel by replacing $\phase{\mu}$ in (\ref{eqn:lemma1_pda}) with $\alpha_0$ to bypass the need for searching the optimum $\mu$, therefore approximating the optimum solution. We formally define the decision rule for EAPQ as
\begin{equation}
\theta_n^{\text{EAPQ}} = \arg \max_{\theta_n\in \Phi_K} \beta^r(\theta_n) \cos(\theta_n + \alpha_n - \alpha_0).
\label{eqn:eapq_decision}
\end{equation}

EAPQ can occasionally select the same discrete phase shifts for some channel realizations, where $e^{j\angle \mu}$ and $e^{j\alpha_0}$ belong to the same arc defined by $e^{j\lambda_l}$ in Algorithm~1. 
However, EAPQ can only guarantee a suboptimal solution, but with a notable complexity reduction compared to Algorithm~1 as the search in the for loop of Algorithm~1 is omitted. 
Yet, EAPQ still requires multiplications for each RIS element, making it more complex than APQ, where only $\alpha_n$ for $n = 0, 1, \ldots, N$ are required for discrete phase shift selection.
\begin{figure*}[!t]
\centering
\begin{minipage}{0.48\textwidth}
\centering
\includegraphics[width=1.1\textwidth]{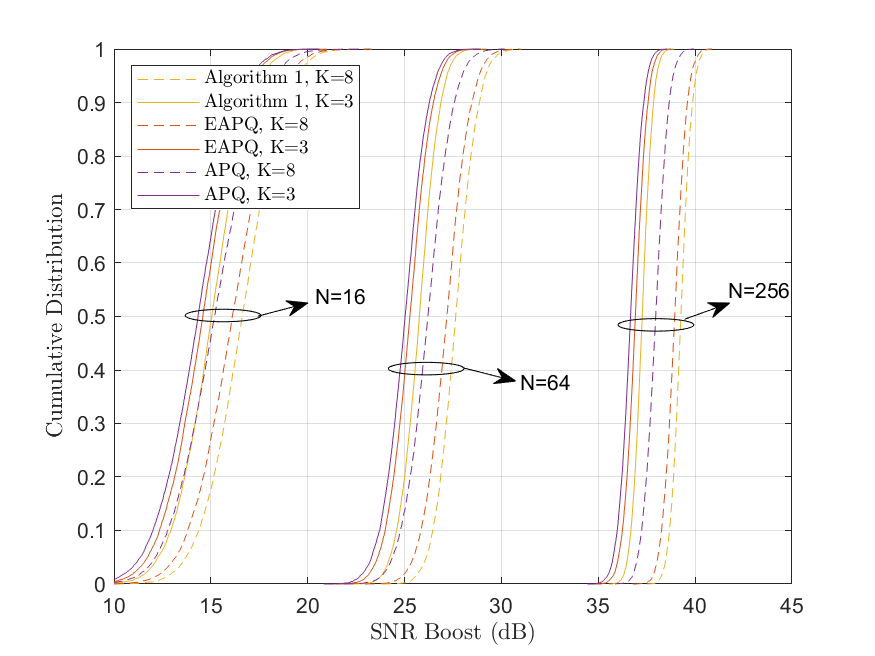}
\caption{CDF plots for SNR Boost with Algorithm~1, EAPQ, and APQ for $\beta^r_{min}=0.2$ and $K\in\{3, 8\}$.}
 \label{fig:CDF_Uniform_02_K3v8}
\end{minipage}%
\hspace{0.03\textwidth}
\begin{minipage}{0.48\textwidth}
\centering
\includegraphics[width=1.1\textwidth]{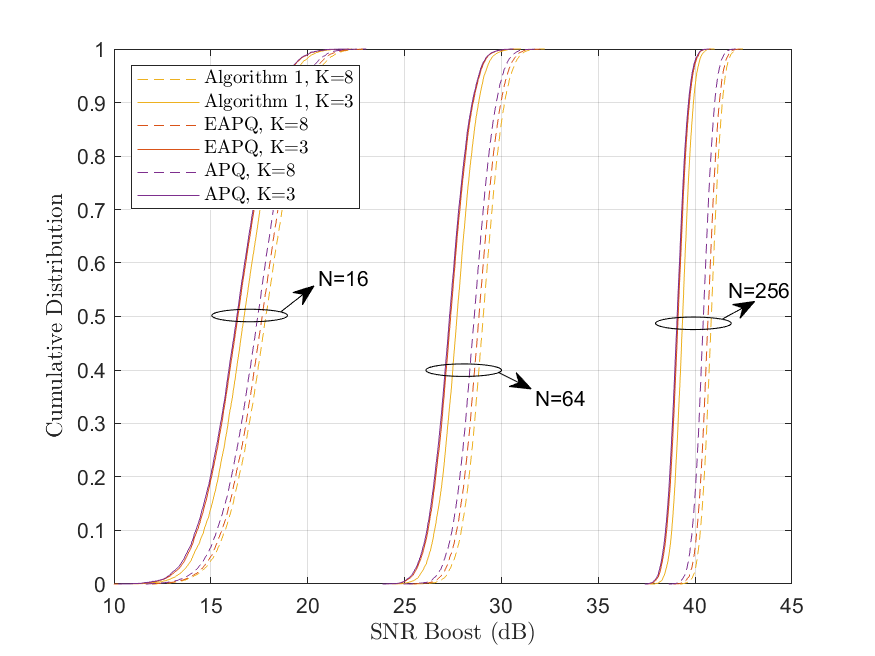}
\caption{CDF plots for SNR Boost with Algorithm~1, EAPQ, and APQ for $\beta^r_{min}=0.5$ and $K\in\{3, 8\}$.}
 \label{fig:CDF_Uniform_05_K3v8}
\end{minipage}
\end{figure*}
\begin{figure*}[!t]
\centering
\begin{minipage}{0.48\textwidth}
\centering
\includegraphics[width=1.0\textwidth]{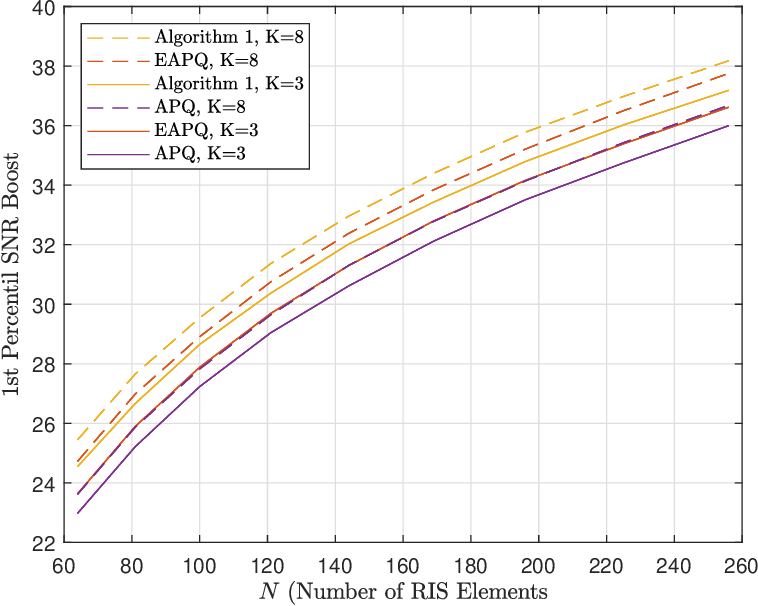}
\caption{1st Percentile SNR Boost vs. $N$, for $\beta^r_{min}=0.2$ and $K\in\{3, 8\}$.}
 \label{fig:Percentile_Uniform_02_K3v8}
\end{minipage}%
\hspace{0.03\textwidth}
\begin{minipage}{0.48\textwidth}
\centering
\includegraphics[width=1.0\textwidth]{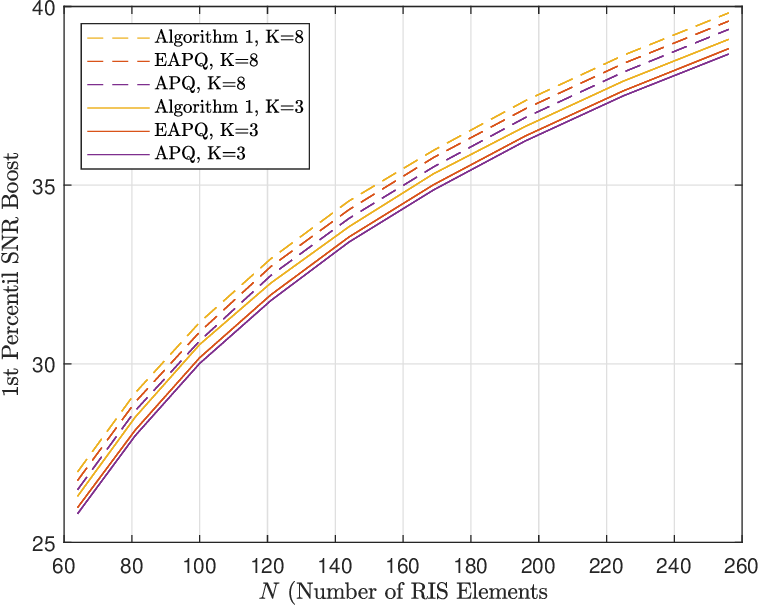}
\caption{1st Percentile SNR Boost vs. $N$, for $\beta^r_{min}=0.5$ and $K\in\{3, 8\}$.}
 \label{fig:Percentile_Uniform_05_K3v8}
\end{minipage}
\end{figure*}

We present the cumulative distribution function (CDF) results for SNR Boost \cite{PA24} in Fig.~\ref{fig:CDF_Uniform_02_K3v8} for $\beta^r_{min}=0.2$, and in Fig.~\ref{fig:CDF_Uniform_05_K3v8} for $\beta^r_{min}=0.5$, where we compare Algorithm~1, APQ, and EAPQ. In these results, we consider $K>2$, to observe the performance gaps between APQ and EAPQ. The CDF results are presented for $N=16,$ $64,$ and $256$ elements, using 10,000 realizations of the channel model defined in \cite{PA24} with $\kappa=0$, where all algorithms ran over the same realization in each step. Between Fig.~\ref{fig:CDF_Uniform_02_K3v8} and Fig.~\ref{fig:CDF_Uniform_05_K3v8}, where the latter has less attenuation, it can be seen that the gain from using larger $K$ and using EAPQ over APQ becomes less. Note that Algorithm~1 with the proven global optimum solution is always superior, even for the worst channel conditions. Finally, although looks marginal, the gain of using Algorithm~1 remains for larger $N$, especially in high attenuation RIS. 
Besides the CDF results, we focus on the low SNR Boost regime and compare the 1st percentile results in Fig.~\ref{fig:Percentile_Uniform_02_K3v8} and Fig.~\ref{fig:Percentile_Uniform_05_K3v8} for $\beta^r_{min}=0.2$ and $\beta^r_{min}=0.5$, respectively. In Fig.~\ref{fig:Percentile_Uniform_02_K3v8}, for larger $K$, EAPQ performs more closely to Algorithm~1. Furthermore, EAPQ with $K=3$ performs the same as APQ with $K=8$, which signifies the superiority of EAPQ under high attenuation. Finaly, with $\beta^r_{min}=0.5$ in Fig.~\ref{fig:Percentile_Uniform_05_K3v8}, the 
performance order is as expected, with performance gaps between the algorithms narrowing in the low SNR Boost regime.

In the next section, for both uniform and nonuniform phases, the achievable performance of the RIS is analyzed under the PDA constraints.

\section{Approximation Ratio of Discrete Phase Shifts and PDA Model with Quantization}
We derived two quantization algorithms, APQ and EAPQ, where the latter is an extended version with the amplitude consideration to perform better under certain scenarios, e.g., for lower $\beta^r_{min}$. Now, we aim to assess the expected performance of the RIS, i.e., the achievable performance. Therefore we derive the approximation ratio for the RIS. 
To that end, we employ the APQ algorithm and its discrete phase shift selections, for the sake of analytical analysis. 
As the numerical results will show, the performance of APQ and EAPQ is relatively close, especially for larger $N$ and mid to high $\beta^r_{min}$, implying that while it is simple to use APQ selections analytically, it can also be a good approximation of the achievable performance in general.

Let us remember the two main non-ideal constraints that are considered: the use of discrete phases that could be either uniform or nonuniform depending on the phase shifting capability of the RIS, i.e., $R$, and the amplitude attenuation that depends on the phase shift selections and $\beta^r_{min}$, namely PDA. The approximation ratio represents the performance of the RIS under non-ideal practical considerations, where it shows how well the ideal solution with ideal considerations can be approximated \cite{PLA24}. Since the channels to be optimized are usually random, we take the ratio of the expected values of the quantization solution and the maximum achievable solution, for large $N$. Therefore, we define the approximation ratio for the PDA model as
\begin{equation}\label{eq:E_PDA_limit}
    E_{PDA} = \lim_{N \to \infty} \frac{\mathbb{E}[f_{\text{rx}}(\boldsymbol{\theta}^\text{APQ})]}{\mathbb{E}[(\sum_{n=0}^{N}\beta_n)^2]}
\end{equation}
where in the denominator, we take the expected value of the maximum achievable result with $\theta_n$. 
To find the numerical value in the numerator, i.e., the expected value of $f_{\text{rx}}(\boldsymbol{\theta}^\text{APQ})$, namely the received power with the APQ solution, we first start by approximating the received power for large $N,$ in a similar way to \cite{PLA24}. Note we make use of (\ref{eq:thetaIdeal}) and state $\theta_n^{\text{ideal}} = \alpha_0 - \alpha_n$ below
\begin{align}
    f_{rx}(\mbox{\boldmath$\theta$}^{\text{APQ}}) =& \bigg|\beta_0e^{j\alpha_0}+\sum_{n=1}^N \beta_n \beta^r(\theta^{\text{APQ}}_n)
	e^{j(\alpha_n + \theta^{\text{APQ}}_n)}\bigg|^2 \nonumber\\
	= & \left|e^{j\alpha_0}\right|^2 \hspace{-0.25em} \bigg|\beta_0 \hspace{-0.25em} + \hspace{-0.25em} \sum_{n=1}^N \beta_n
	\beta^r(\theta^{\text{APQ}}_n) e^{j(\alpha_n+\theta^{\text{APQ}}_n-\alpha_0)}\bigg|^2 \nonumber\\
	= & \bigg|\beta_0+\sum_{n=1}^N \beta_n
	\beta^r(\theta^{\text{APQ}}_n) e^{j(\theta^{\text{APQ}}_n-\theta^{\text{ideal}}_n)}\bigg|^2\nonumber\\
	\approx & \bigg|\sum_{n=1}^N \beta_n
	\beta^r(\theta^{\text{APQ}}_n) e^{j(\theta^{\text{APQ}}_n-\theta^{\text{ideal}}_n)}\bigg|^2. \label{eqn:frx}
\end{align}
Let $\delta_n = \theta^{\text{APQ}}_n-\theta^{\text{ideal}}_n$ for $n = 1, 2, \dots, N.$ The resulting term above can be rewritten as \cite{PLA24}\
\ifCLASSOPTIONonecolumn
\begin{align}    
    f_{rx}(\boldsymbol{\theta}^{\text{APQ}}) &\approx \bigg|\sum_{n=1}^N\beta_n \beta^r(\theta^{\text{APQ}}_n) e^{j(\theta^{\text{APQ}}_n-\theta^{\text{ideal}}_n)}\bigg|^2 \nonumber \\
    &= \sum_{n=1}^N \beta_n^2 \nonumber \\
    &\,\,\,\,\,\,\,\,\,\,\,+ \; 2 \sum_{k=2}^N\sum_{l=1}^{k-1} \beta_k\beta_l \beta^r(\theta^{\text{APQ}}_k)\beta^r(\theta^{\text{APQ}}_l)\cos(\delta_k - \delta_l).
    \label{eqn:frxApxCos}
\end{align}
Therefore
\begin{equation}
    \mathbb{E}[f_{rx}(\boldsymbol{\theta}^{\text{APQ}})] = N\mathbb{E}[\beta_n^2] + N(N-1)\mathbb{E}[\beta_k\beta_l]\mathbb{E}[\beta^r(\theta^{\text{APQ}}_k)\beta^r(\theta^{\text{APQ}}_l)\cos(\delta_k - \delta_l)]. \label{eq:E_PDA_nominator}
\end{equation}
\else
\small
\begin{align}    
    f_{rx}&(\boldsymbol{\theta}^{\text{APQ}}) \nonumber \\
    &\approx \bigg|\sum_{n=1}^N\beta_n \beta^r(\theta^{\text{APQ}}_n) e^{j(\theta^{\text{APQ}}_n-\theta^{\text{ideal}}_n)}\bigg|^2 \nonumber \\
    &= \sum_{n=1}^N \beta_n^2+ \; 2 \sum_{k=2}^N\sum_{l=1}^{k-1} \beta_k\beta_l \beta^r(\theta^{\text{APQ}}_k)\beta^r(\theta^{\text{APQ}}_l)\cos(\delta_k - \delta_l).
    \label{eqn:frxApxCos}
\end{align}
Therefore
\begin{align}
    \mathbb{E}[f_{rx}(\boldsymbol{\theta}^{\text{APQ}})&] = N\mathbb{E}[\beta_n^2] \nonumber \\ 
    &\hspace{-2.5em}+ N(N-1)\mathbb{E}[\beta_k\beta_l]\mathbb{E}[\beta^r(\theta^{\text{APQ}}_k)\beta^r(\theta^{\text{APQ}}_l)\cos(\delta_k - \delta_l)]. \label{eq:E_PDA_nominator}
\end{align}
\normalsize
\fi
Since $\mathbb{E}[(\sum_{n=1}^{N}\beta_n)^2] = N\mathbb{E}[\beta_n^2] + N(N-1)\mathbb{E}[\beta_k\beta_l]$, for $N \gg 1$, the resulting approximation ratio will be the ratio of the terms with $N^2$ in (\ref{eq:E_PDA_limit}), resulting in
\begin{align}\label{eq:E_PDA_step1}
    E_{PDA} &= \frac{\mathbb{E}[\beta_k\beta_l]\mathbb{E}[\beta^r(\theta^{\text{APQ}}_k)\beta^r(\theta^{\text{APQ}}_l)\cos(\delta_k - \delta_l)]}{\mathbb{E}[\beta_k\beta_l]} \nonumber \\
    &= \mathbb{E}[\beta^r(\theta^{\text{APQ}}_k)\beta^r(\theta^{\text{APQ}}_l)\cos(\delta_k - \delta_l)].
\end{align}

As the final step, we assume all $\beta_k, \beta_l, \delta_k$, and $\delta_l$ are independent from each other. With a trigonometric identity on the cosine term, $E_{PDA}$ can be further simplified as
\small
\begin{equation}\label{eq:E_PDA_step2}
    E_{PDA} \hspace{-0.25em} = \hspace{-0.25em}\left(\mathbb{E}\left[\beta^r(\theta^{\text{APQ}}_n)\cos\left(\delta_n\right)\right]\right)^2 \hspace{-0.65em}+ \hspace{-0.25em}\left(\mathbb{E}\left[\beta^r(\theta^{\text{APQ}}_n)\sin\left(\delta_n\right)\right]\right)^2.
\end{equation}\normalsize

Due to the symmetric structure of the discrete phase shifts and the PDA curve on the complex plane, the second term with the sine function in (\ref{eq:E_PDA_step2}) will be zero \cite{PLA24}. Therefore, it is sufficient to calculate $\mathbb{E}\left[\beta^r(\theta^{\text{APQ}}_n)\cos\left(\delta_n\right)\right]$ and take the square of the final result. With the total law of expectation, we have
\begin{align}\label{eq:totalEx_1}
    \mathbb{E} & \left[\beta^r(\theta^{\text{APQ}}_n)\cos\left(\delta_n\right)\right] = \nonumber \\
    &\sum_{k=1}^{K} \left[ p(\phi_k) \beta^r(\phi_k) \int_{-\pi}^{\pi} f(\delta_n|\theta_n = \phi_k) \cos(\delta_n) d\delta_n\right],
\end{align}
where we took $\beta^r(\phi_k)$ outside the integral. The general form of the approximation ratio is the square of (\ref{eq:totalEx_1}). 
Using this form in the following subsections, we will calculate the closed form solutions for $E_{PDA}$ under two main considerations: first for uniform discrete phases with sufficiently large $R$, second, with nonuniform discrete phase shifts for $R < 2\pi\frac{K-1}{K}$.

\subsection{Approximation with PDA Model for Uniform Discrete Phase Shifts}

In (\ref{eq:totalEx_1}), the gain term $\beta^r(\phi_k)$ is deterministic for given RIS coefficients $\beta^r_{min}$, $\alpha^r$, and $\phi^r$, where it is calculated with the PDA model given in (\ref{eq:PDAmodel}). 
Note that our algorithms will be such that they will work for any arbitrary ${\bf W}_K$, i.e., any arbitrary phase shift and gain pair that does not necessarily comply with the model in (\ref{eq:PDAmodel}). In (\ref{eq:totalEx_1}), $p(\phi_k)$ represents the probability of selecting $\phi_k$. Therefore, for uniform discrete phases, $p(\phi_k)=1/K$ for $k=1,\dots,K$ as $\theta_n^{\text{ideal}}$ are i.i.d. uniform. Furthermore, for a given $\phi_k$, $\delta_n \sim {\cal U}[-\frac{\pi}{K}, \frac{\pi}{K}]$ due to the uniformly placed discrete phase shifts, which gives $f(\delta_n|\theta_n = \phi_k) = K/2\pi$ for $k=1,\dots,K$. Hence, for uniform discrete phase shifts, we update (\ref{eq:totalEx_1}) as
\begin{align}
    \mathbb{E} & \left[\beta^r(\theta^{\text{APQ}}_n)\cos\left(\delta_n\right)\right] \nonumber \\
    &= \left[\frac{1}{K}\sum_{k=1}^{K} \beta^r(\phi_k)\right]\left[ \frac{K}{2\pi}\int_{-\pi/K}^{\pi/K} \cos(\delta_n) d\delta_n\right] \label{eq:totalEx_uniform0} \\
    &= \frac{\sinc(1/K)}{K} \sum_{k=1}^{K} \beta^r(\phi_k). \label{eq:totalEx_uniform1}
\end{align}

Therefore, the approximation ratio for the uniform discrete phases, that is, $E_{PDA}(K)$, will be the square of (\ref{eq:totalEx_uniform1}) as given below:
\begin{equation}
	E_{PDA}(K) = \left(\frac{\sinc(1/K)}{K} \sum_{k=1}^{K} \beta^r(\phi_k)\right)^2. \\
\end{equation}

\subsubsection{Observations on the Approximation Ratio with Uniform Discrete Phases}

We make two observations on the result in (\ref{eq:totalEx_uniform0}). First, note that the values inside the square brackets correspond to the individual expectations of the RIS gains and the quantization error, i.e., $\mathbb{E}[\beta^r(\theta_n)]\mathbb{E}[\cos\left(\delta_n\right)]$. Therefore, we see that the two are actually uncorrelated, and each contribute to the performance loss separately. However, we do remark that for the practical RIS model that we consider in this paper, $\mathbb{E}[\beta^r(\theta_n)]$ depends on the number of quantization levels as it determines the points we sample the PDA curve. But, for an arbitrary ${\bf W}_K$, there may not be any dependence between them. Nevertheless, we can define the loss compared to the ideal scenario in dB as
\begin{align}
    L_{dB}^{\rm uniform} &= 10\log_{10} (\mathbb{E}[\beta^r(\theta_n)]^2 E[\cos(\delta_n)]^2) \nonumber \\
    &= 20\left(\log_{10}(\mathbb{E}[\beta^r(\theta_n)]) + \log_{10}(\mathbb{E}[\cos\left(\delta_n\right)])\right) \label{eq:lossindb},
\end{align}
where, in the following section with nonuniform discrete phases, we will not see a similar independent effect on the performance loss.

In Table~\ref{tbl:LossdBTable}, we present the loss in dB for $\beta^r_{min} \in \{0.2, 0.5, 0.8\}$ and $K\in \{2, 3, 4, 6, 8\}$.
It can be seen that, the gain of changing the attenuation level from high to mid, i.e., $\beta^r_{min}=0.2$ to $\beta^r_{min}=0.5$, is about $2$ dB for $K>2$. Besides this, the gain of changing the attenuation level from mid to low is about $1.5$ dB for $K>3$ and increases with lower quantization levels. This shows the increased sensitivity to lower quantization levels, for lower attenuation levels.
On the other hand, no matter what the attenuation level is, using more than $K=4$ discrete phase shifts provides marginal gains.
\begin{table}[!t]
	\begin{center}\small
		\begin{tabular}{lllccccc}
			\hline
			&$K\!=\!2$&$K\!=\!3$&$K\!=\!4$&$K\!=\!6$&$K\!=\!8$\\
			\hline
			\hline
			$\beta^r_{min} = 0.2$&8.359&7.252&6.395&5.906&5.731\\
			\hline
			$\beta^r_{min} = 0.5$&6.421&4.712&3.918&3.416&3.242\\
			\hline
			$\beta^r_{min} = 0.8$&4.838&2.749&1.993&1.485&1.309\\
			\hline
		\end{tabular}
		\caption{Loss (in dB) compared to ideal scenario, calculated by equation (\ref{eq:lossindb}) for $\beta^r_{min} \in \{0.8, 0.5, 0.2\}$ and $K\in \{2, 3, 4, 6, 8\}$.}
		\label{tbl:LossdBTable}
	\end{center}
\end{table}

\subsubsection{Observation on the Approximation Ratio for Continuous Phase Shifts with Amplitude Attenuation}

Finally, we let $K \rightarrow \infty$. With uniform discrete phases, this would correspond to $\theta_n^{\text{ideal}} \in [-\pi, \pi)$ making the error term $\delta_n = 0$ and $\mathbb{E}[\cos(\delta_n)] = 1$, resulting in $E_{PDA}^{\text{cont.}}=\mathbb{E}[\beta^r(\theta_n)]$. In this case, the expectation $\mathbb{E}[\beta^r(\theta_n)]$ will be the integral of the PDA curve for $\theta_n \in [-\pi, \pi),$ resulting in
\begin{equation}\label{eq:approx_pda_continuous}
    E_{PDA}^{\rm cont.} = \left(\frac{1}{2\pi} \int_{0}^{2\pi}\beta^r(\theta)d\theta\right)^2.
\end{equation}

The result in (\ref{eq:approx_pda_continuous}) is compatible with the analysis and the result provided in \cite{AZWY20}, where the loss for a single-user point-to-point scenario is determined solely by the attenuation of the RIS elements. In other words, when the discrete phases are relaxed, the approximation to the ideal case will be the expected value of the gains of the RIS element, which is the mean of the PDA curve given by (\ref{eq:PDAmodel}).
\begin{figure*}[!t]
\centering
\begin{minipage}{0.48\textwidth}
\centering
\includegraphics[width=1.0\textwidth]{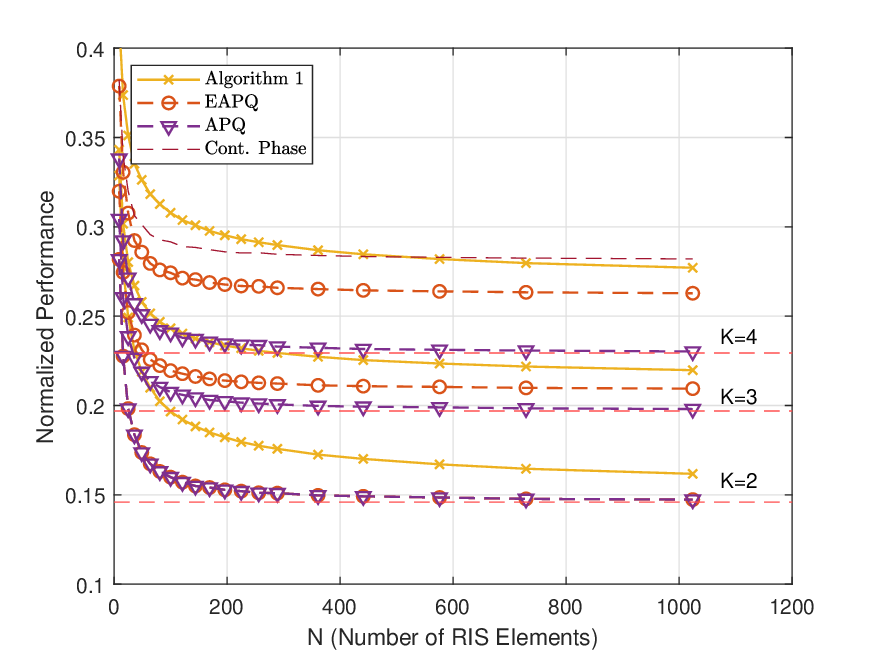}
\caption{Normalized Performance results vs. $N$, for $\beta^r_{min}=0.2$ and $K\in\{2,3,4\}$.}
 \label{fig:approx_uniform_bmin02}
\end{minipage}%
\hspace{0.03\textwidth}
\begin{minipage}{0.48\textwidth}
\centering
\includegraphics[width=1.0\textwidth]{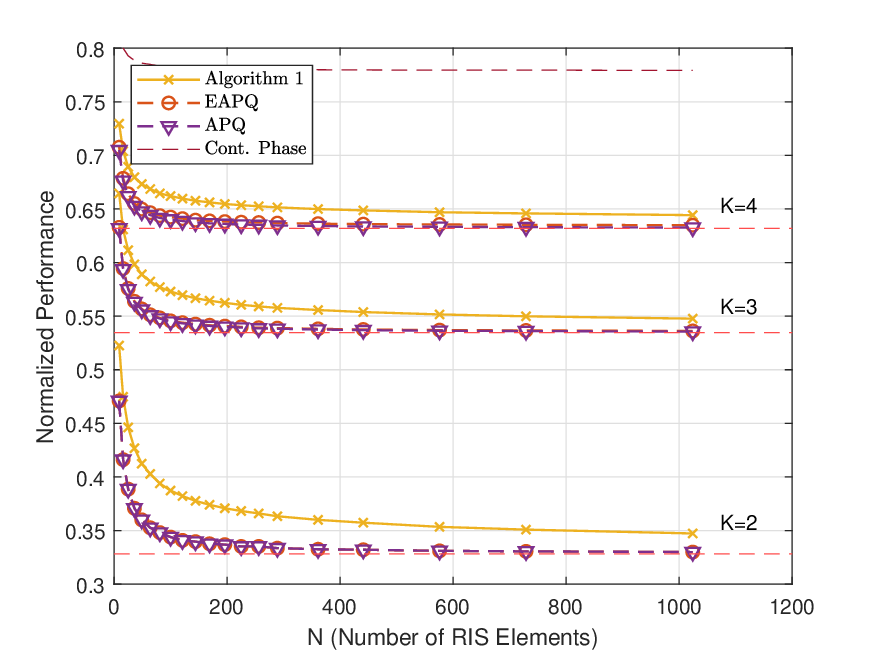}
\caption{Normalized Performance results vs. $N$, for $\beta^r_{min}=0.8$ and $K\in\{2,3,4\}$.}
 \label{fig:approx_uniform_bmin08}
\end{minipage}
\end{figure*}

We present the simulation results for the normalized performance with uniform discrete phases in Fig.~\ref{fig:approx_uniform_bmin02} for $\beta^r_{min}=0.2$, and in Fig.~\ref{fig:approx_uniform_bmin08} for $\beta^r_{min}=0.8$. In these results, Algorithm~1, APQ, and EAPQ ran over the same channel realizations, where the result of each algorithm was divided by $(\sum_{n=0}^{N}\beta_n)^2$ for normalization. In both figures, our theoretical result $E_{PDA}(K)$ for uniform discrete phases successfully approximates the APQ result for large $N$, therefore validating our calculations. When there is significant gain attenuation, i.e., in Fig.~\ref{fig:approx_uniform_bmin02} with $\beta^r_{min}=0.2$, the performance gap between EAPQ and APQ increases with larger $K$. 
Furthermore, another interesting observation is that, Algorithm~1 can outperform the continuous solution for up to $N = 500$ elements. This is because the continuous solution is defined based on the ideal assumption and it is not the optimum when there is gain attenuation. From Fig.~\ref{fig:approx_uniform_bmin02} to Fig.~\ref{fig:approx_uniform_bmin08}, as $\beta^r_{min}$ grows larger, the performance gap between APQ and EAPQ decreases, showing that the amplitude-aware quantization technique becomes less crucial with lower attenuation. In general, it can be seen that the global optimum solution provided by Algorithm~1 with discrete phase shifts is always superior to the quantization solution.

Next, we will consider a larger limitation in the RIS phase range, i.e. $R<2\pi\frac{K-1}{K}$, and provide further analysis and information on the approximation ratio with non-uniform discrete phase shifts.

\subsection{Approximation with PDA Model with a Limited RIS Phase Range}

In this subsection, we consider nonuniform discrete phase shifts with a limited phase range as shown in Fig.~\ref{fig:DPS_nonuniform}. Unlike the uniform case, the probability mass function for selecting $\phi_k$ will not be constant. For $K>2$, $p(\phi_k)$ is defined as
\begin{equation}\label{eq:pmfNonuniform}
p(\phi_k) = 
\left\{
\begin{array}{ll}
\frac{A}{2\pi}  & \,\, \text{for}\,\,\,\, \phi_1 \\
\frac{R}{2\pi(K-1)}  & \,\, \text{for}\,\,\,\, \phi_2 \\
& \vdots \\
\frac{R}{2\pi(K-1)}  & \,\, \text{for}\,\,\,\, \phi_{K-1} \\
\frac{A}{2\pi}  & \,\, \text{for}\,\,\,\, \phi_K \\
\end{array}
\right.
\end{equation}
where $A = \left(\pi - \frac{R}{2} + \frac{R}{2(K-1)}\right)$. 
Note that, each value in $p(\phi_k)$ is divided by $2\pi$ due to the uniform distribution assumption on the ideal solution.
When $K = 2$, the value of $A$ becomes $\pi$, and the probability mass function $p(\phi_k)$ becomes uniform, since the angular range over which the ideal solution is quantized to either of the two phase options spans an equal angle of $\pi$.
Let us assume $K>2$ in (\ref{eq:totalEx_1}), or equivalently $K\geq3$, and calculate the square root of the approximation ratio as
\begin{align}\label{eq:totalEx_nonuniform}
    \mathbb{E} & \left[\beta^r(\theta^{\text{APQ}}_n)\cos\left(\delta_n\right)\right] \nonumber \\
    =&\left[\frac{A}{2\pi}\beta^r(\phi_1) + \frac{A}{2\pi}\beta^r(\phi_K)\right]\int_{-\frac{R}{2(K-1)}}^{\pi-\frac{R}{2}}\frac{1}{A}\cos(\delta_n)d\delta_n \nonumber \\
    &+\sum_{k=2}^{K-1} \frac{R}{2\pi(K-1)}\beta^r(\phi_k) \int_{-\frac{R}{2(K-1)}}^{\frac{R}{2(K-1)}}\frac{1}{\frac{R}{(K-1)}}\cos(\delta_n)d\delta_n \nonumber \\
    =&\frac{\beta^r(\phi_1) + \beta^r(\phi_K)}{2\pi} \left(\sin\left(\frac{R}{2(K-1)}\right) + \sin\left(\frac{R}{2}\right) \right) \nonumber \\
    &\hspace{1em}+\frac{1}{\pi}\left[\sum_{k=2}^{K-1} \beta^r(\phi_k)\right] \sin\left(\frac{R}{2(K-1)}\right).
\end{align}
Note that, if we assumed $K=2$, we would not have the second term with the summation in (\ref{eq:totalEx_nonuniform}), where there would only be $\phi_1$ and $\phi_{K=2}$. Therefore, let us define the approximation ratio for the nonuniform case for two different scenarios:
\begin{align}
    &E_{PDA} (R, K) = \frac{1}{\pi^2} \Bigg(\left[\sum_{k=2}^{K-1} \beta^r(\phi_k)\right] \sin\left(\frac{R}{2(K-1)}\right)\nonumber \\
    & + \hspace{-0.25em}\frac{\beta^r(\phi_1) + \beta^r(\phi_K)}{2}\hspace{-0.25em} \left(\sin\left(\frac{R}{2(K-1)}\right) \hspace{-0.25em}+ \sin\left(\frac{R}{2}\right) \hspace{-0.25em}\right) \hspace{-0.25em}\Bigg)^2, \label{eq:E_PDA_nonunif_Kgt2}
\end{align}
where $K>2$. For $K=2$:
\begin{equation}
	E_{PDA}(R, K=2) = \frac{\sin^2(R/2)}{\pi^2}\left(\beta^r(\phi_1) + \beta^r(\phi_K)\right)^2. \label{eq:E_PDA_nonunif_K2}
\end{equation}

\begin{figure*}[!t]
\centering
\begin{minipage}{0.48\textwidth}
\centering
\includegraphics[width=1.0\textwidth]{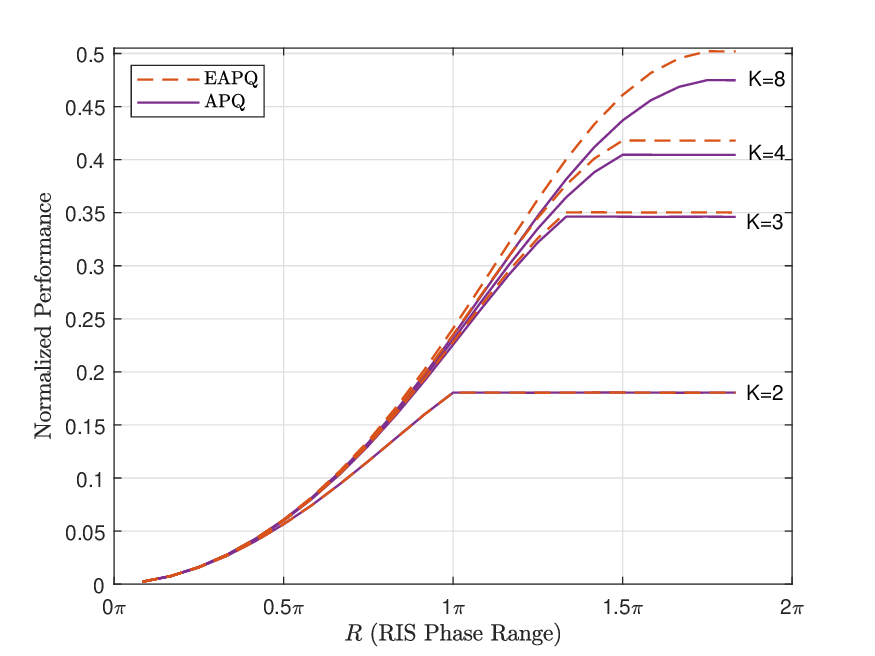}
\caption{Normalized Performance results vs. $R$, for $\beta^r_{min}=0.5$, $N=1024$ and $K\in\{2,3,4,8\}$.}
 \label{fig:approx_nonuniform_empiric}
\end{minipage}%
\hspace{0.03\textwidth}
\begin{minipage}{0.48\textwidth}
\centering
\includegraphics[width=1.0\textwidth]{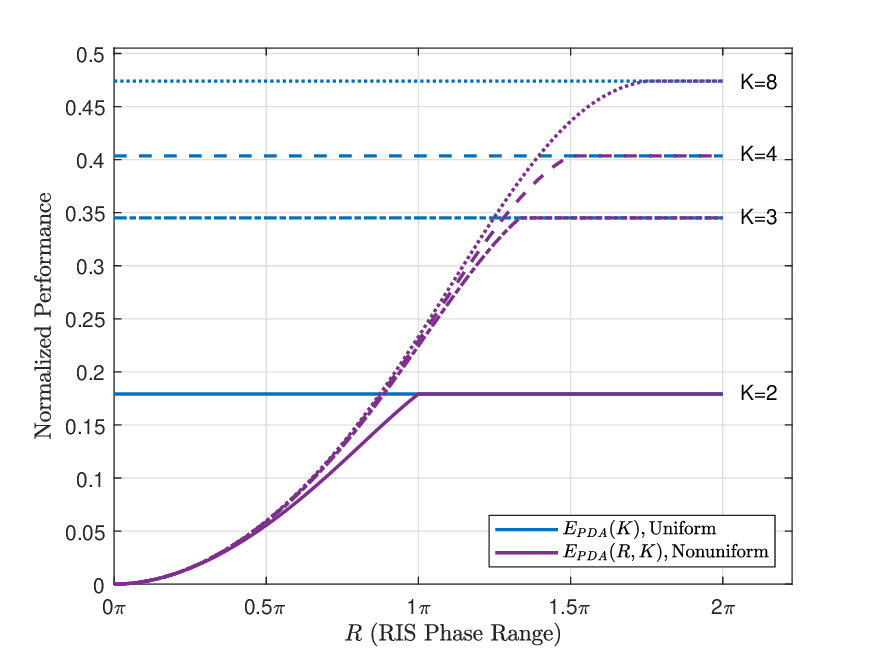}
\caption{$E_{PDA}(R,K)$ vs. $R$, for $\beta^r_{min}=0.5$ and $K\in\{2,3,4,8\}$.}
 \label{fig:approx_nonuniform_theoric}
\end{minipage}
\end{figure*}

We present the normalized performance results for both EAPQ and APQ under moderate gain attenuation, e.g., $\beta^r_{min}=0.5$ in Fig.~\ref{fig:approx_nonuniform_empiric} using simulation results, and for APQ in Fig.~\ref{fig:approx_nonuniform_theoric} using the approximation ratio. In Fig.~\ref{fig:approx_nonuniform_empiric}, as the RIS phase increases, EAPQ can bring more performance compared to APQ, especially for larger $K$. 
An illustration of the approximation $E_{PDA}(R,K)$ is given in Fig. \ref{fig:approx_nonuniform_theoric} together with $E_{PDA}(K)$, where it can be seen that $E_{PDA}(R,K)$ converges to the approximation ratio of the uniform phases, as the RIS phase range increases. 
Furthermore, comparing Fig.~\ref{fig:approx_nonuniform_empiric} and Fig.~\ref{fig:approx_nonuniform_theoric}, the empirical and theoretical performance of APQ are similar, again validating the approximation ratio for the nonuniform discrete phases.
\begin{figure}[!t]
\centering
\includegraphics[width=0.50\textwidth]{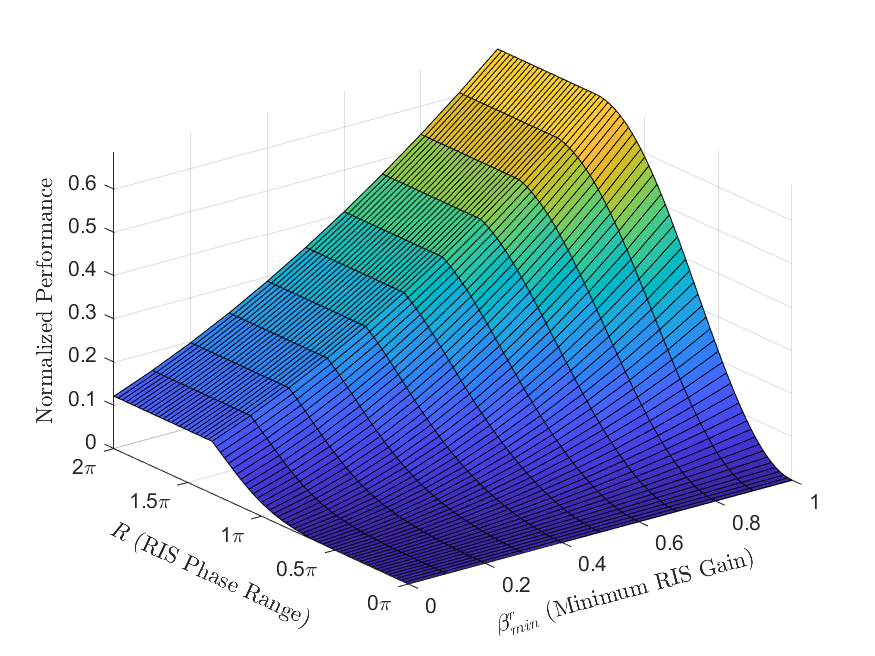}
\caption{Normalized Performance results vs. $R \in [0, 2\pi]$ and $\beta^r_{min} \in [0, 1]$, for $K=3$.}
 \label{fig:approx_surf}
\end{figure}

Finally, as an overall view of the joint impact of the PDA model and the RIS phase range, we present the normalized performance result in Fig.~\ref{fig:approx_surf} for $K=3$, where the normalized performance for each possible $\beta^r_{min}$ and $R$ pair is shown. The values are calculated with the closed-form solution $E_{PDA}(R, K=3)$. 
For sufficient RIS phase range, i.e., $R \geq 2\pi\frac{K-1}{K} = 240^\circ$ with $K=3$, the performance remains constant for constant $\beta^r_{min}$ as the uniform discrete phase shift structure in Fig.~\ref{fig:DPS_uniform} remains the same. It is also important to note that given a limited phase range, the performance becomes more sensitive to the attenuation parameter $\beta^r_{min}$ for larger $R$. In addition to this, the performance becomes more sensitive to $R$ for less attenuation with larger $\beta^r_{min}$.
\section{Convergence to Optimality and Complexity}
In this section, we briefly analyze the complexity of Algorithm~1 and compare it with the optimal algorithms proposed in our earlier works. In Algorithm~1, the main for-loop (Steps 4–11) consists of $\sum_{l=1}^{L} \mathcal{O}(|\mathcal{N}(\lambda_l)|) = \mathcal{O}(NK)$ operations, where each step involves two vector additions to update a single element. Additionally, the initialization phase (Steps 2–3), which computes the initial coefficients and stores the corresponding gain using {\em Lemma~1}, requires $N$ vector additions. As a result, the total computational cost of Algorithm~1 amounts to $N(2K+1)$ vector additions.

Table~\ref{tab:complexityComparison_PDA} presents a comparison of Algorithm~1 with the optimal algorithms from \cite{PA24} and \cite{PLA24}, which handle the uniform and nonuniform discrete phase shift cases, respectively. When the gains of RIS coefficients are independent of the phase, i.e., constant gain, the complexity reduces to $N(K+1)$ vector additions, while in the adjustable gain case it becomes $N(K+2)$, as reported in \cite{PLA24}. Furthermore, for strictly uniform discrete phase shifts, the complexity can be further reduced to $N$ vector additions \cite{PA24}.

It is worth noting that the higher complexity of Algorithm~1 is expected, as the previous algorithms are special cases of the more general setting considered in this work. Without Algorithm~1, solving the $K$-ary quadratic programming (QP) problem under the PDA constraint would require exponential complexity.

\begin{table}[!t]\small
\centering
\caption{Comparison of Algorithm~1 with the optimum algorithms in earlier works.}
\begin{tabular}{|c|c|c|c|}
                \hline
                & Search Steps & Time  & RIS\\
                &  			& Complexity & Coefficients\\
                \hline\hline
Opt.	& $\leq N$ & $\mathcal{O}(N)$ & Const. Gain\\
\cite{PA24}     &  &  & Uniform \\
                \hline
Opt.-1   & $\leq NK$ & $\mathcal{O}(N(K+1))$ & Const. Gain\\
\cite{PLA24}    &  &  & Nonuniform \\
                \hline
Opt.-2   & $\leq N(K+1)$ & $\mathcal{O}(N(K+2))$ & On/Off Gain\\
\cite{PLA24}    &  &  & Nonuniform \\
                \hline
                \hline
\textbf{Algo.}	& $\boldsymbol{\leq NK}$ & $\boldsymbol{\mathcal{O}(N(2K+1))}$ & PDA\\
\textbf{1}&  &  &  \\
                \hline
\end{tabular}
\label{tab:complexityComparison_PDA}\small
\end{table}

\section{Conclusion}
In this work, we investigated the received power maximization problem for RIS-assisted communications under phase-dependent amplitude (PDA) constraints and discrete phase shifts that are within a limited RIS phase range. We established the necessary and sufficient conditions to achieve optimality in convex amplitude-phase configurations and proposed a linear-time search algorithm that guarantees convergence within $NK$ steps. 
Additionally, we introduced the APQ and EAPQ quantization frameworks as low complexity algorithms that work surprisingly well under certain scenarios. Furthermore, with the quantization approach, we analyzed the approximation ratio to quantify the expected performance from an RIS with practical limitation, compared to the ideal scenario with continuous phases and no attenuation. Our analysis demonstrated that, for a sufficient RIS phase range, increasing the phase resolution beyond $K=4$ provides limited performance benefits, and that the system is more sensitive to amplitude attenuation when the phase range is large, and vice versa. On the other hand, when the RIS phase range is limited, we showed that the performance is more sensitive to attenuation for larger phase range, and more sensitive to phase range for less attenuation. Finally, our proposed optimal algorithm, i.e., Algorithm~1, can serve as a versatile upper bound and a benchmark for discrete RIS beamforming under realistic hardware constraints. Therefore, this work complements existing studies on received power maximization by addressing discrete beamforming under practical hardware constraints, such as phase-dependent amplitude attenuation. In that regard, it provides an extension to our prior works \cite{PA24, PLA24}.

\bibliographystyle{IEEEtran}
\bibliography{ref}
\end{document}